\documentclass[11pt,a4paper]{article}
\pdfoutput=1
\usepackage{jheppub}
\usepackage{amsthm,amsbsy,amsfonts,mathrsfs,enumerate,float,wrapfig,amsmath}
\usepackage[utf8]{inputenc}
\usepackage[outline]{contour}
\DeclareUnicodeCharacter{2212}{-}
\usepackage{subfigure}

\newcommand{\be}{\begin{equation}}
\newcommand{\ee}{\end{equation}}
\newcommand{\bea}{\begin{eqnarray}}
\newcommand{\eea}{\end{eqnarray}}
\newcommand{\ba}{\begin{align}}
\newcommand{\ea}{\end{align}}

\newcommand{\CBHS}[3]{\footnotesize{$\dfrac{#1}{#2}$ \newline $\lvert \rvert$ \newline $\begin{array}{l} #3 +\ldots \end{array}$}}
\newcommand{\HBHS}[3]{\footnotesize{$\dfrac{#1}{#2} = \begin{array}{l} #3 +\ldots \end{array}$}}
\usepackage{longtable,caption,multirow}
\usepackage{amsmath}
\usepackage{amssymb}
\usepackage{MnSymbol}
\DeclareMathOperator{\PE}{PE}
\DeclareMathOperator{\HWG}{HWG}

\DeclareMathOperator{\MQ}{MQ}

\usepackage{array}
\newcolumntype{L}[1]{>{\raggedright\let\newline\\\arraybackslash\hspace{0pt}}m{#1}}
\newcolumntype{C}[1]{>{\centering\let\newline\\\arraybackslash\hspace{0pt}}m{#1}}
\newcolumntype{R}[1]{>{\raggedleft\let\newline\\\arraybackslash\hspace{0pt}}m{#1}}

\usepackage{tikz}
\usetikzlibrary{decorations.pathmorphing}
\usetikzlibrary{decorations.pathreplacing}
\usetikzlibrary{shapes, shapes.geometric, shapes.symbols, shapes.arrows, shapes.multipart, shapes.callouts, shapes.misc}
\tikzset{snake it/.style={decorate, decoration=snake}}
\tikzset{7brane/.style={circle, draw=black, fill=black,ultra thick,inner sep=1 pt, minimum size=1 pt,}, c/.default={4pt}}

\tikzset{big7brane/.style={circle, draw=black, fill=black,ultra thick,inner sep=2.5 pt, minimum size=1 pt,}, c/.default={4pt}}
\tikzset{u/.style={circle, draw=black, fill=white, thick,inner sep=2 pt, minimum size=2 pt,},f/.style={square, draw=black, fill=white,ultra thick,inner sep=4 pt, minimum size=2 pt,}}

\tikzset{so/.style={circle, draw=black, fill=red, thick,inner sep=2 pt, minimum size=2 pt,},f/.style={square, draw=black, fill=white,ultra thick,inner sep=4 pt, minimum size=2 pt,}}

\tikzset{sp/.style={circle, draw=black, fill=blue,thick,inner sep=2 pt, minimum size=2 pt,},f/.style={square, draw=black, fill=white,ultra thick,inner sep=4 pt, minimum size=2 pt,}}

\tikzset{uf/.style={rectangle, draw=black, fill=white, thick,inner sep=2.5 pt, minimum size=4 pt,}}

\tikzset{spf/.style={rectangle, draw=black, fill=blue, thick,inner sep=2.5 pt, minimum size=4 pt,}}

\tikzset{sof/.style={rectangle, draw=black, fill=red, thick,inner sep=2.5 pt, minimum size=4 pt,}}
\usetikzlibrary{positioning}
\usetikzlibrary{arrows}

\title{Exploring the orthosymplectic zoo}
\author[a]{Mohammad Akhond,}
\author[b]{Federico Carta,}
\author[c]{Siddharth Dwivedi,}
\author[d]{Hirotaka Hayashi,}
\author[e]{Sung-Soo Kim,}
\author[f]{and Futoshi Yagi}
\affiliation[a]{Department of Physics, Swansea University, \\
 Singleton Park, Swansea, SA2 8PP, U.K.}
\affiliation[b]{Department of Mathematical Sciences, Durham University,\\
	Durham, DH$1$ $3$LE, United Kingdom}
\affiliation[c]{Center for Theoretical Physics, College of Physical Science and Technology, Sichuan University, \\Chengdu, 610064, China}
\affiliation[d]{Department of Physics, School of Science, Tokai University,\\ 4-1-1 Kitakaname, Hiratsuka-shi, Kanagawa 259-1292, Japan}
\affiliation[e]{School of Physics, University of Electronic Science and Technology of China, \\
No.2006, Xiyuan Ave, West Hi-Tech Zone, 
Chengdu, Sichuan 611731, China}
\affiliation[f]{School of Mathematics, Southwest Jiaotong University,\\ 
West zone, High-tech district, Chengdu, Sichuan 611756, China}
\emailAdd{akhondmohammad@gmail.com}
\emailAdd{federico.carta@durham.ac.uk}
\emailAdd{sdwivedi@scu.edu.cn}
\emailAdd{h.hayashi@tokai.ac.jp}
\emailAdd{sungsoo.kim@uestc.edu.cn}
\emailAdd{futoshi\_yagi@swjtu.edu.cn}

\abstract{We study the Higgs branch of the SCFT limit of 5d SO(6) and SO(8) gauge theory with hypermultiplets in the spinor and vector representations. In the case of SO(6) gauge theories, we contrast the magnetic quivers obtained with those of SU(4) gauge theory with hypermultiplets in the fundamental and second rank antisymmetric representations. Since SU(4) gauge theories admit several different values of the Chern-Simons level, we make some observations about how to distinguish those theories from the brane webs of the SO(6) theories. In the case of SO(8) gauge theories, we use SO(8) triality to propose (naively) inequivalent magnetic quivers, which will turn out to have the same moduli spaces of vacua, at least locally around their most singular loci. We encounter several interesting new phenomena occurring in the magnetic quivers, such as hypermultiplets between neighbouring symplectic gauge nodes and matter in two-index representations of unitary nodes. We also give a prescription for computing the local Coulomb branch Hilbert series for quivers involving bad USp(2) gauge nodes.}
\begin{document}
\maketitle
\section{Introduction}
Superconformal field theories (SCFTs) in five spacetime dimensions have received much deserved attention in recent years. Their mere existence forces us to expand our view of quantum field theories, as these fixed points cannot be reached by traditional means of perturbing around free field Lagrangians. Yet, by now, there is overwhelming evidence to support their existence, mostly due to stringy constructions starting with the seminal papers \cite{Seiberg:1996bd,Intriligator:1997pq,Ganor:1996pc,Brandhuber:1999np}. Broadly speaking, there are three independent, yet complementary points of view for studying these SCFTs, namely their embedding into type IIB brane webs \cite{Aharony:1997bh,Aharony:1997ju,Bergman:2013ala,Bergman:2013aca,Bergman:2014kza,Bergman:2015dpa,Kim:2015jba,Zafrir:2015ftn,Hayashi:2015vhy,Zafrir:2015rga,Hayashi:2015zka,Zafrir:2016jpu,Hayashi:2016abm,Hayashi:2016jak,Hayashi:2017jze,Hayashi:2017btw,Hayashi:2018bkd,Hayashi:2018lyv,Hayashi:2019yxj,Kim:2020hhh,Kim:2021fxx}, geometric engineering \cite{Douglas:1996xp,Morrison:1996xf,Intriligator:1997pq,Jefferson:2018irk,Closset:2018bjz,Bhardwaj:2019jtr,Apruzzi:2019kgb,Bhardwaj:2020gyu,Tian:2021cif,Acharya:2021jsp} and holography \cite{ Bergman:2018hin,Fluder:2018chf,Kaidi:2018zkx,Uhlemann:2019lge,Uhlemann:2019ypp,Legramandi:2021uds,Legramandi:2021aqv}.

Many 5d SCFTs, admit supersymmetry preserving mass deformations, which trigger an RG flow, whose low energy dynamics is effectively captured by an $\mathcal{N}=1$ gauge theory. Such deformations, preserve the SU(2)$_R$ symmetry, while breaking the flavour symmetry. An important dynamical question is therefore to determine the full global symmetry of the parent SCFT of a given 5d gauge theory. 5d SCFTs and gauge theories, can possess a Higgs branch of their moduli space of vacua, which in the gauge theory regime can be constructed as the hyperK\"ahler quotient \cite{Gray:2008yu,Hanany:2008kn}. In the SCFT limit, the hyperK\"ahler quotient, is no longer accessible, due to a lack of a Lagrangian description, making the study of the Higgs branch in this limit more challenging.
There are, by now, a plethora of techniques to determine the enhanced global symmetry of the SCFT parent of a given 5d gauge theory, such as 7-brane analysis \cite{Sen:1996vd,Gaberdiel:1997ud,Gaberdiel:1998mv,DeWolfe:1998eu,DeWolfe:1998pr,DeWolfe:1999hj}, superconformal indices \cite{Kim:2012gu,Bergman:2013koa, Hwang:2014uwa}, as well as geometric approaches \cite{Katz:1996xe,Witten:1996qb,Morrison:1996xf,Douglas:1996xp,Intriligator:1997pq, Jefferson:2018irk,Apruzzi:2019opn,Apruzzi:2019enx,Bhardwaj:2020ruf,Bhardwaj:2020avz}. One particularly elegant approach to determine the SCFT flavour symmetries as well as the Higgs branch of the SCFT, pioneered in \cite{Ferlito:2017xdq}, is to consider their magnetic quivers. The magnetic quiver (MQ), of a given 5d theory, is a 3d $\mathcal{N}=4$ quiver gauge theory, whose Coulomb branch is isomorphic to the Higgs branch of the 5d theory in question. In many cases, though not always, one can show that the magnetic quiver of a given 5d theory, is the 3d mirror of its torus compactification. This leads to an interesting interplay between 5d $\mathcal{N}=1$ theories and 3d $\mathcal{N}=4$ theories, and has prompted many recent studies \cite{Akhond:2020vhc,Akhond:2021knl,Akhond:2021ffo,vanBeest:2020civ,vanBeest:2020kou,vanBeest:2021xyt,Bourget:2019rtl,Bourget:2020asf,Bourget:2020gzi,Bourget:2020mez,Bourget:2020xdz,Bourget:2021xex,Closset:2020afy,Closset:2020scj,Cabrera:2019izd,Ferlito:2017xdq,Carta:2021whq,Carta:2021dyx}. 

That 3d magnetic quivers are advantageous, is due to recent advances in extracting the algebraic geometry of 3d $\mathcal{N}=4$ Coulomb branches \cite{Cremonesi:2013lqa,Cremonesi:2014kwa,Cremonesi:2014vla,Cremonesi:2014xha,Hanany:2014dia}. Coulomb branches of 3d $\mathcal{N}=4$ theories, are parameterised by BPS monopole operators, whose R-charges can be determined from the R-charges of the fermionic content of the  \cite{Borokhov:2002cg}. Following the terminology of Gaiotto and Witten \cite{Gaiotto:2008ak}, we refer to a 3d $\mathcal{N}=4$ theory as good, ugly or bad, depending on the conformal dimension of the monopole operators in that theory. A theory is said to be good, if the conformal dimension of all monopole operators is greater than $\frac{1}{2}$, which is the dimension of a scalar in (2+1) dimensions. A bad theory is one in which some monopole operator has a conformal dimension less than $\frac{1}{2}$. Finally, in an ugly theory, there are some monopole operators whose conformal dimension is exactly $\frac{1}{2}$, but none that are smaller. For good or ugly theories, the monopole formula \cite{Cremonesi:2013lqa}, can correctly produce the Hilbert series of the Coulomb branch, using the UV gauge theory data. However, for bad theories, the monopole formula fails, due to the fact that the UV R-symmetries of the theory are different from the superconformal R-symmetry, and hence the UV R-charge cannot be used to predict the conformal dimension of the SCFT operators. The full structure of the Coulomb branch of some bad theories was recently understood by Assel and Cremonesi in \cite{Assel:2017jgo,Assel:2018exy}, building on earlier results in \cite{Yaakov:2013fza,Ferlito:2016grh}. In particular, it was found that for USp$(2N)$ SQCD with $2N$ flavours in the fundamental representation, the Coulomb branch has two most singular points, where the theory flows to an interacting SCFT in the deep infra-red, with a local mirror Lagrangian in the vicinity of each singular point. This is unlike the situation for good theories, where the most singular point on the moduli space is unique, namely the origin, with an SCFT at the bottom of the RG flow in that vacuum. The Coulomb branch Hilbert series of the local geometry around one of the two most singular points of the USp($2N$) SQCD with $2N$ flavours was computed in \cite{Assel:2018exy}, and shown to agree with the Higgs branch Hilbert series of the local mirror.

The goal of this paper is to illuminate the Higgs branch of the SCFT parents of 5d gauge theories whose gauge group is either SO(6), or SO(8), and with matter in the vector, spinor, and conjugate spinor representations. Our motivation is partly due to the fact that, among classical simple gauge groups, the magnetic quivers for $A_N$ and $C_N$ cases are well studied \cite{Cabrera:2018jxt,Ferlito:2017xdq}, while those of $B_N$ and $C_N$ remain relatively unexplored. The magnetic quivers for SO$(N)$ gauge theories with matter in vector representation were recently constructed in \cite{Akhond:2021ffo}, using their embedding into brane webs with O7$^+$-planes. The current work complements this study by adding matter in the spinor and the conjugate spinor representations. The reason for restricting the rank of the gauge group is mostly for practical reasons. The construction of magnetic quivers from brane webs with O5-planes, is at present, still not completely systematic. It is, therefore, useful to limit the discussion to situations where a consistency check of our computations is available. For SO(6) gauge theories, this is achieved by comparing the orthosymplectic magnetic quivers obtained from brane webs with an O5-plane, with the unitary magnetic quivers obtained from the brane webs of SU(4) gauge theories. Similarly, for SO(8) gauge theories, one can set up a consistency check, by exploiting SO(8) triality. As we shall see, in the rest of this paper, such considerations also lead to some interesting results for 3d $\mathcal{N}=4$ theories. In particular, in Section \ref{SO6 theories}, we exploit the isomorphism between SO(6) and SU(4), to conjecture exact highest weight generating functions for orthosymplectic magnetic quivers, simply by carrying over the results from their unitary counterparts. Similarly, in Section \ref{SO8 theories}, we uncover several intriguing equalities of moduli spaces, of naively unrelated quivers.

Throughout the paper, we encounter magnetic quivers which contain bad USp$(2N)$ nodes, where the effective number of hypermultiplets is exactly $2N$. We devise a  method to associate a Hilbert series to these quivers, by using the local mirror description around one of their two most singular loci. We will refer to this procedure of associating the Coulomb branch Hilbert series of a good theory, to the local geometry of the Coulomb branch near a singular locus of a bad theory as  ``B2G". For the specific case of USp$(2)$ with 2 flavours, the prescription is first encountered in section \ref{sec:SO6+1S1C1V}. A similar prescription for USp($2N$) theories for $N>2$ is used throughout the paper, though the technical details will be published elsewhere \cite{HSbad}. The validity of our prescription is confirmed, by comparing the results with those computed using the Hall-Littlewood technique \cite{Cremonesi:2014kwa}. In addition, many of the magnetic quivers that involve bad nodes, can have a good ``dual" description, since we have several inequivalent brane constructions for each 5d theory we consider. The Hilbert series computed using our proposed prescription, is also consistent with the good ``dual" quivers. 

The content of the remainder of this paper is organised as follows: Section \ref{SO6 theories} contains MQs for SO(6) gauge theories derived from brane webs with O5-planes, as well as MQs for SU(4) theories, derived from ordinary brane webs. Here we will encounter MQs with bad symplectic gauge nodes and devise various methods to extract the good interacting part. Section \ref{SO8 theories} is dedicated to studying the MQs for SO(8) theories, and a comparison is made between MQs of SO(8) theories related by triality, exchanging spinor, conjugate spinor, and vector representations of SO(8). We conclude with our wish list for future projects in \ref{conclusions}. Appendix \ref{appendixA} contains the unrefined Hilbert series results for the OSp quivers.
\input{SO6}
\section{SO(8) triality}\label{SO8 theories}
In this section we consider SO(8) gauge theories with matter in the vector \textbf{V}, spinor \textbf{S} and conjugate spinor \textbf{C} representations. We then use SO(8) triality to produce several equivalent magnetic quivers for a given theory. We will encounter theories whose Higgs branch is given as the union of several cones, where each cone is described by a distinct magnetic quiver. We will denote by $\MQ_{i;\, \text{SO}(8)}^{s,c,v}$, the magnetic quiver for the $i$-th cone of the SCFT parent of SO(8) +$s$\textbf{S}+$c$\textbf{C}+$v$\textbf{V}, i.e.
\begin{equation}
    \mathcal{H}^{5\text{d}}_\infty\left(\text{SO}(8)+s\textbf{S}+c\textbf{C}+v\textbf{V}\right)=\bigcup_{i}\mathcal{C}^{3\text{d}}\left(\MQ_{i;\, \text{SO}(8)}^{s,c,v}\right) \ .
\end{equation}
\subsection{SO(8)+1\textbf{V} \texorpdfstring{$\longleftrightarrow$}{TEXT} SO(8)+1\textbf{S}}
The web diagram for SO(8)+1\textbf{V} at infinite coupling is given by
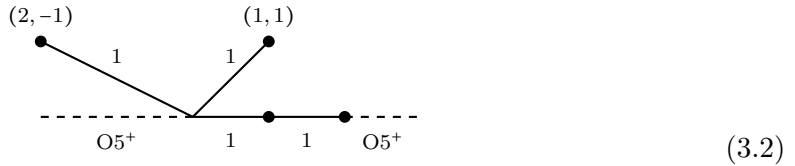
\begin{equation}
\begin{scriptsize}
    \begin{tikzpicture}
             \draw[thick,dashed](0,0)--(2,0);
    \draw[thick](4,0)--(2,0);
    \draw[thick,dashed](4,0)--(5,0);
    \draw[thick](2,0)--(3,1);
    \draw[thick](2,0)--(0,1);
    \node[7brane]at(4,0){};
    \node[7brane]at(3,0){};
    \node[label=above:{$(2,-1)$}][7brane]at(0,1){};
    \node[label=above:{$(1,1)$}][7brane]at(3,1){};
    \node[label=below:{O5$^+$}]at(1,0){};
    \node[label=below:{O5$^+$}]at(4.5,0){};
    \node[label=below:{$1$}]at(3.5,0){};
    \node[label=below:{$1$}]at(2.5,0){};
    \node[label=above:{$1$}]at(1,.5){};
    \node[label=above:{$1$}]at(2.5,.5){};
    \end{tikzpicture}
\end{scriptsize}
\end{equation}
The magnetic quiver that we obtain from here is given by
\begin{equation}
\MQ_{1;\, \text{SO}(8)}^{0,0,1} = \begin{array}{c}
     \begin{scriptsize}
        \begin{tikzpicture}
                 \node[label=below:{1}][so](o1){};
                 \node[label=below:{2}][sp](sp2)[left of=o1]{};
                 \node[label=below:{1}][u](u1)[left of=sp2]{};
                 \node[label=below:{1}][uf](uf1)[left of=u1]{};
                 \node[label=above:{3}][uf](uf2)[above of=u1]{};
                 \node[label=above:{1}][sof](sof)[above of=sp2]{};
                 \draw(o1)--(sp2);
                 \draw(sp2)--(u1);
                 \draw(u1)--(uf2);
                 \draw(sp2)--(sof);
                 \path [draw,snake it](u1)--(uf1);
        \end{tikzpicture}
    \end{scriptsize}
\end{array}  \quad \contour{black}{$\xrightarrow{\text{B2G}}$} \quad \begin{array}{c}
         \begin{scriptsize}
             \begin{tikzpicture}
                 \node[label=below:{1}][u](u11){};
                 \node[label=below:{1}][u](u1)[left of=u11]{};
                 \node[label=below:{1}][uf](uf1)[left of=u1]{};
                 \node[label=above:{3}][uf](uf2)[above of=u1]{};
                 \node[label=above:{1}][uf](uf3)[above of=u11]{};
                 \draw(u11)--(u1);
                 \draw(u1)--(uf2);
                 \draw(u11)--(uf3);
                 \path [draw,snake it](u1)--(uf1);
             \end{tikzpicture}
         \end{scriptsize}
    \end{array}\;.
\label{SO8:MQ001}
\end{equation}
On the other hand, the brane web for the infinite coupling limit of SO(8)+1\textbf{S} is given by
\begin{equation}
    \begin{array}{c}
         \begin{scriptsize}
             \begin{tikzpicture}
                      \draw[thick,dashed](3,0)--(-2,0);
                      \draw[thick](0,0)--(-2,1);
                      \draw[thick](0,0)--(3,1.5);
                      \node[label=above:{$(2,-1)$}][7brane]at(-2,1){};
                      \node[label=above:{$(2,1)$}][7brane]at(3,1.5){};
                      \node[7brane]at(2,1){};
                      \node[label=below:{O5$^+$}]at(-1,0){};
                      \node[label=below:{O5$^-$}]at(1.5,0){};
                      \node[label=above:{1}]at(-1,.5){};
                      \node[label=above:{2}]at(1,.5){};
                      \node[label=above:{1}]at(2,1){};
             \end{tikzpicture}
         \end{scriptsize}
    \end{array}\;,
\end{equation}
whose magnetic quiver is given by
\begin{equation}
   \MQ_{1;\, \text{SO}(8)}^{1,0,0} = \begin{array}{c}
         \begin{scriptsize}
             \begin{tikzpicture}
                 \node[label=below:{1}][u](u11){};
                 \node[label=below:{1}][u](u1)[left of=u11]{};
                 \node[label=below:{1}][uf](uf1)[left of=u1]{};
                 \node[label=above:{3}][uf](uf2)[above of=u1]{};
                 \node[label=above:{1}][uf](uf3)[above of=u11]{};
                 \draw(u11)--(u1);
                 \draw(u1)--(uf2);
                 \draw(u11)--(uf3);
                 \path [draw,snake it](u1)--(uf1);
             \end{tikzpicture}
         \end{scriptsize}
    \end{array}\;.
    \label{SO8:MQ100}
\end{equation}
This coincides with the magnetic quiver obtained in \cite{Akhond:2021knl} from the brane web with an O7$^+$-plane. The unrefined Coulomb and Higgs branch Hilbert series of the two quivers agree and are given in table \ref{CoulombHSOSp} and table \ref{HiggsHSOSp} respectively. We further propose the following highest weight generating function of the Coulomb branch: 
\begin{equation}
    \HWG_{\mathcal{C}}(\ref{SO8:MQ001})=\HWG_{\mathcal{C}}(\ref{SO8:MQ100})=\PE\left[\left(1+\mu^2\right)t^2+\left(q+q^{-1}\right)\mu t^6-\mu^2t^{12}\right] ~,
\end{equation}
where $\mu$ is the SU(2) fugacity and $q$ is the U(1) charge.
\subsection{SO(8)+2\textbf{V} \texorpdfstring{$\longleftrightarrow$}{TEXT} SO(8)+2\textbf{S}}
The web diagram for SO(8)+2\textbf{V} at infinite coupling is given by
\begin{equation}
    \begin{array}{c}
         \begin{scriptsize}
             \begin{tikzpicture}
                      \draw[thick,dashed](0,0)--(1,0);
                      \draw[thick](1,0)--(5,0);
                      \draw[thick,dashed](5,0)--(6,0);
                      \draw[thick](3,0)--(4,1);
                      \draw[thick](3,0)--(2,1);
                      \node[7brane] at (1,0){};
                      \node[7brane] at (2,0){};
                      \node[7brane] at (4,0){};
                      \node[7brane] at (5,0){};
                      \node[label=above:{$(1,-1)$}][7brane] at (2,1){};
                      \node[label=above:{$(1,1)$}][7brane] at (4,1){};
                      \node[label=below:{O5$^+$}]at(0.5,0){};
                      \node[label=below:{O5$^+$}]at(5.5,0){};
                      \node[label=below:{$1$}]at(1.5,0){};
                      \node[label=below:{$1$}]at(2.5,0){};
                      \node[label=below:{$1$}]at(3.5,0){};
                      \node[label=below:{$1$}]at(4.5,0){};
             \end{tikzpicture}
         \end{scriptsize}
    \end{array}\;,
\end{equation}
and the corresponding magnetic quiver is given by
\begin{equation}
    \MQ_{1;\, \text{SO}(8)}^{0,0,2} = \begin{array}{c}
         \begin{scriptsize}
             \begin{tikzpicture}
                      \node[label=below:{1}][so](o1){};
                      \node[label=below:{2}][sp](sp1)[right of=o1]{};
                      \node[label=below:{3}][so](so3)[right of=sp1]{};
                      \node[label=below:{2}][sp](sp1')[right of=so3]{};
                      \node[label=below:{1}][so](o1')[right of=sp1']{};
                      \node[label=left:{1}][u](u1)[above of=so3]{};
                      \node[label=above:{3}][uf](uf)[above of=u1]{};
                      \draw(o1)--(sp1);
                      \draw(sp1)--(so3);
                      \draw(so3)--(sp1');
                      \draw(sp1')--(o1');
                      \draw(so3)--(u1);
                      \draw(u1)--(uf);
             \end{tikzpicture}
         \end{scriptsize}
    \end{array}\;.
    \label{SO8:MQ002}
\end{equation}
By SO(8) triality, the 5d theory SO(8)+2\textbf{V} should be identical to SO(8)+2\textbf{S}, the latter has a web at infinite coupling given by
\begin{equation}
    \begin{array}{c}
         \begin{scriptsize}
             \begin{tikzpicture}
                      \draw[thick,dashed](-1,0)--(5,0);
                      \draw[thick](2,0)--(5,1.5);
                      \draw[thick](2,0)--(-1,1.5);
                      \node[label=above:{(2,1)}][7brane]at (5,1.5){};
                      \node[label=above:{(2,-1)}][7brane]at (-1,1.5){};
                      \node[7brane]at (4,1){};
                      \node[7brane]at (0,1){};
                      \node[label=below:{O5$^-$}]at (-.5,0){};
                      \node[label=below:{O5$^-$}]at (4.5,0){};
                      \node[label=above left:{2}]at(3,.5){};
                      \node[label=above right:{2}]at(1,.5){};
                      \node[label=above left:{1}]at(4.5,1.25){};
                      \node[label=above right:{1}]at(-.5,1.25){};
             \end{tikzpicture}
         \end{scriptsize}
    \end{array}\;.
\end{equation}
Here, we encounter a 5-brane web diagram that includes more than one coincident subwebs that intersect with O5-plane.
In this case, we need to consider the two types of contribution:
One is the contribution coming from one of the subweb and its mirror image, which corresponds to the factor $|2m_i|$ in the monopole formula. The other is the contribution coming from the mirror pair of different subwebs among coincident ones, which corresponds to the factor $|m_i + m_j|$ $(i < j)$ in the monopole formula. As discussed in \cite{Akhond:2020vhc}, the number of each contribution is given by
\begin{align}
\# \text{ of } |2m_i|&:~~ \frac{1}{2} (\text{Self SI}) - (\text{SI with O5} )\ ,
\cr
\# \text{ of } |m_i + m_j| &:~~   (\text{Self SI})  \qquad (i < j)\ .
\end{align}
Here, ``SI'' denotes the stable intersection number, which is used to compute the number of hypermultiplets, and ``Self SI'' means the SI with the mirror images of the considered subwebs. 

Both types of contributions need to be part of the weight of representations of the unitary group. Such weights are included in two representations: 
One is in rank 2 symmetric tensor representation, which we denote as ``Sym$^2$'', and the other is in rank 2 antisymmetric tensor, which we denote as ``$\wedge^2$''.
The weight $|2m_i|$ is included only in Sym$^2$ while the weight $ |m_i + m_j| $ is included in both of the representations. 
In order to reproduce the contribution mentioned above, we claim that the number of these hypermultiplets are given by
\begin{align}
\# \text{ of Sym$^2$}:\quad \frac{1}{2} (\text{Self SI}) - (\text{SI with O5} )\ ,
\cr
\# \text{ of } \wedge^2:\quad  \frac{1}{2} (\text{Self SI}) + (\text{SI with O5} )\ .
\end{align}
Note that ``Sym$^2$'' is interpreted as the charge 2 hypermultiplets when the gauge group is U(1). Indeed, the proposal for Sym$^2$ above is the generalization of the rule proposed in \cite{Akhond:2020vhc} regarding the number of the charge 2 hypermultiplets. Also, it was proposed in \cite{Akhond:2020vhc} that the number of $\wedge^2$ is simply given by Self SI, which now turned out to be valid only when the number of Sym$^2$ is 0. For generic case, we need to modify the rule as mentioned above. 

If we apply this proposal to coincident $(p,q)$ 5-branes intersecting with O5$^-$-plane, we have
\begin{align}
\# \text{ of Sym$^2$}:\quad pq - q\ ,
\cr
\# \text{ of } \wedge^2:\quad  pq + q\ .
\end{align}
By using this, we claim that the magnetic quiver for the SCFT limit of SO(8)+2\textbf{S} is given by
\begin{equation}
    \MQ_{1;\, \text{SO}(8)}^{2,0,0} = \begin{array}{c}
         \begin{scriptsize}
             \begin{tikzpicture}
                      \node[label=below:{1}][u](1){};
                      \node[label=above left:{2}][u](2)[right of=1]{};
                      \node[label=below:{1}][u](1')[right of=2]{};
                      \node[label=above:{1}][uf](uf)[above of=2]{};
                      \draw(1)--(2);
                      \draw(2)--(1');
                      \path [draw,snake it](2)--node[right] {$\;\text{Sym}^2$}++(uf);
                      \draw (2)to[out=-45,in=225,loop,looseness=10](2);
                      \node[label=above:{$3\;\wedge^2$}][below of=2]{};
             \end{tikzpicture}
         \end{scriptsize}
    \end{array} ~.
    \label{SO8:MQ200}
    \end{equation}
The unrefined Coulomb and Higgs branch Hilbert series of \eqref{SO8:MQ002} and \eqref{SO8:MQ200} match and are given in table \ref{CoulombHSOSp} and table \ref{HiggsHSOSp} respectively. We further propose the following highest weight generating function of the Coulomb branch: 
\begin{equation}
    \HWG_{\mathcal{C}}(\ref{SO8:MQ002})=\HWG_{\mathcal{C}}(\ref{SO8:MQ200})=\PE\left[\left(1+\mu_1^2\right)t^2+\mu_2^2t^4+\left(q+q^{-1}\right)\mu_2 t^6-\mu_2^2t^{12}\right] ~,
\end{equation}
where $\mu_i$ are highest weight fugacities for USp$(4)$, and $q$ is the fugacity for U(1).
\subsection{SO(8)+2\textbf{S}+1\textbf{C} \texorpdfstring{$\longleftrightarrow$}{TEXT} SO(8)+2\textbf{S}+1\textbf{V} \texorpdfstring{$\longleftrightarrow$}{TEXT} SO(8)+1\textbf{S}+2\textbf{V}}
The brane web for SO(8)+2\textbf{S}+1\textbf{C} at infinite coupling is given by
\begin{equation}
\begin{array}{c}
     \begin{scriptsize}
         \begin{tikzpicture}
                  \draw[thick,dashed](-1,0)--(2,0);
                  \draw[thick](2,0)--(-1,1.5);
                  \node[label=above:{(2,-1)}][7brane]at (-1,1.5){};
                  \node[7brane]at (.5,.75){};
                  \draw[thick](2,0)--(3,1);
                  \node[label=above:{(1,1)}][7brane]at (3,1){};
                  \draw[thick](2,0)--(6,0);
                  \node[7brane]at (4,0){};
                  \node[7brane]at (3,0){};
                  \node[7brane]at (5,0){};
                  \node[7brane]at (6,0){};
                  \draw[thick,dashed](7,0)--(6,0);
                  \node[label=below:{O5$^-$}]at(-.5,0){};
                  \node[label=below:{O5$^-$}]at(6.5,0){};
                  \node[label=below:{2}]at(2.5,0){};
                  \node[label=below:{$\frac{3}{2}$}]at(3.5,0){};
                  \node[label=below:{$1$}]at(4.5,0){};
                  \node[label=below:{$\frac{1}{2}$}]at(5.5,0){};
                  \node[label=above:{1}]at(0,1){};
                  \node[label=above:{2}]at(1,.5){};
                  \node[label=above:{2}]at(2.25,.5){};
         \end{tikzpicture}
     \end{scriptsize}
\end{array}
\end{equation}
from which we read off the magnetic quiver
\begin{equation}
    \MQ_{1;\, \text{SO}(8)}^{2,1,0} = \begin{array}{c}
         \begin{scriptsize}
             \begin{tikzpicture}
                      \node[label=below:{2}][so](so2){};
                      \node[label=below:{2}][sp](sp2)[right of=so2]{};
                      \node[label=above left:{2}][u](u2)[right of=sp2]{};
                      \node[label=below:{1}][u](u1)[right of=u2]{};
                      \node[label=above:{1}][uf](uf)[above of =u2]{};
                      \draw(so2)--(sp2);
                      \draw(sp2)--(u2);
                      \draw[dashed](u1)--(u2);
                      \path [draw,snake it](u2)--node[right] {$\;\text{Sym}^2$}++(uf);
                      \draw (u2)to[out=-45,in=225,loop,looseness=10](u2);
                      \node[label=above:{$2\wedge^2$}][below of=u2]{};
             \end{tikzpicture}
         \end{scriptsize}
    \end{array} ~.
    \label{SO8:MQ210}
\end{equation}
The brane web for SO(8)+2\textbf{S}+1\textbf{V} at infinite coupling is given by
\begin{equation}
\begin{array}{c}
     \begin{scriptsize}
         \begin{tikzpicture}
                  \draw[thick,dashed](-1,0)--(2,0);
                  \draw[thick](2,0)--(-1,1.5);
                  \node[label=above:{(2,-1)}][7brane]at (-1,1.5){};
                  \node[7brane]at (.5,.75){};
                  \draw[thick](2,0)--(4,2);
                  \node[label=above:{(1,1)}][7brane]at (4,2){};
                  \draw[thick](2,0)--(4,0);
                  \node[7brane]at (4,0){};
                  \node[7brane]at (3,1){};
                  \node[7brane]at (3,0){};
                  \draw[thick,dashed](4,0)--(5,0);
                  \node[label=below:{O5$^-$}]at(-.5,0){};
                  \node[label=below:{O5$^-$}]at(4.5,0){};
                  \node[label=below:{2}]at(2.5,0){};
                  \node[label=below:{$\frac{3}{2}$}]at(3.5,0){};
                  \node[label=above:{1}]at(0,1){};
                  \node[label=above:{2}]at(1,.5){};
                  \node[label=above:{2}]at(2.25,.5){};
                  \node[label=above:{2}]at(3.25,1.5){};
         \end{tikzpicture}
     \end{scriptsize}
\end{array}
\end{equation}
From here we obtain the following magnetic quiver
\begin{equation}
    \MQ_{1;\, \text{SO}(8)}^{2,0,1} = \begin{array}{c}
         \begin{scriptsize}
             \begin{tikzpicture}
                      \node[label=below:{1}][u](u1){};
                      \node[label=left:{2}][u](u2)[above right of=u1]{};
                      \node[label=below:{1}][u](u1')[right of=u2]{};
                      \node[label=above:{2}][sp](sp2)[above left  of=u2]{};
                      \node[label=above:{1}][uf](uf)[above of=u2]{};
                      \draw(u1)--(u2);
                      \draw(u2)--(u1');
                      \draw(u2)--(sp2);
                      \path [draw,snake it](u2)--node[right] {$\;\text{Sym}^2$}++(uf);
                      \draw (u2)to[out=-45,in=225,loop,looseness=10](u2);
                      \node[label=above:{$2\wedge^2$}][below of=u2]{};
             \end{tikzpicture}
         \end{scriptsize}
    \end{array} \quad \contour{black}{$\xrightarrow{\text{B2G}}$} \quad 
    \begin{array}{c}
         \begin{scriptsize}
             \begin{tikzpicture}
                      \node[label=below:{1}][u](u1){};
                      \node[label=left:{2}][u](u2)[above right of=u1]{};
                      \node[label=below:{1}][u](u1')[right of=u2]{};
                      \node[label=above:{1}][u](sp2)[above left  of=u2]{};
                      \node[label=above:{1}][uf](uf)[above of=u2]{};
                      \draw(u1)--(u2);
                      \draw(u2)--(u1');
                      \draw[dashed](u2)--node[left]{$\frac{1}{2}$}(sp2);
                      \path [draw,snake it](u2)--node[right] {$\;\text{Sym}^2$}++(uf);
                      \draw (u2)to[out=-45,in=225,loop,looseness=10](u2);
                      \node[label=above:{$2\wedge^2$}][below of=u2]{};
             \end{tikzpicture}
         \end{scriptsize}
    \end{array} ~,
    \label{SO8:MQ201}
\end{equation}
where the contribution of the dashed line to the monopole formula is given by (see footnote \ref{footnote1})
\begin{equation}
    \Delta_\text{hyp}\left(\begin{array}{c}
         \begin{scriptsize}
        \begin{tikzpicture}
                 \node[label=below:{1}][u](1){};
                 \node[label=below:{2}][u](2)[right of=1]{};
                 \draw[dashed](1)--node[above]{$\frac{1}{2}$}++(2);
        \end{tikzpicture}
    \end{scriptsize}\end{array}\right)=\frac{1}{2}\left(|m_1-m_{2,1}|+|m_1+m_{2,2}|\right)\;,
\end{equation}
where $m_1$ is the magnetic flux of the U(1) node, while $m_{2,i}$ are magnetic fluxes for the U(2) gauge node. 
The brane web for SO(8)+1\textbf{S}+2\textbf{V} is given by
\begin{equation}
    \begin{array}{c}
         \begin{scriptsize}
             \begin{tikzpicture}
                  \draw[thick, dashed](2,0)--(3,0);
                      \draw[thick, dashed](7,0)--(8,0);
                      \draw[thick](3,0)--(7,0);
                      \node[7brane]at(3,0){};
                      \node[7brane]at(4,0){};
                      \node[7brane]at(6,0){};
                      \node[7brane]at(7,0){};
                      \node[7brane]at(6,1){};
                      \draw[thick](5,0)--(7,2);
                      \draw[thick](5,0)--(4,1);
                      \node[label=above:{$(1,1)$}][7brane]at(7,2){};
                      \node[label=above:{$(1,-1)$}][7brane]at(4,1){};
                      \node[label=below:{O5$^+$}]at(2.5,0){};
                      \node[label=below:{$1$}]at(3.5,0){};
                      \node[label=below:{$2$}]at(5.5,0){};
                      \node[label=below:{$2$}]at(4.5,0){};
                      \node[label=below:{$\frac{3}{2}$}]at(6.5,0){};
                      \node[label=below:{O5$^-$}]at(7.5,0){};
                      \node[label=above:{1}]at(6.5,1.5){};
                      \node[label=above:{2}]at(5.5,.5){};
                      \node[label=above:{1}]at(4.5,.5){};
             \end{tikzpicture}
         \end{scriptsize}
    \end{array}\;.
\end{equation}
From here we read off the following magnetic quiver:
\begin{equation}
    \MQ_{1;\, \text{SO}(8)}^{1,0,2} = \begin{array}{c}
         \begin{scriptsize}
             \begin{tikzpicture}
                      \node[label=below:{1}][so](o1){};
                      \node[label=below:{2}][sp](sp2)[right of=o1]{};
                      \node[label=below:{3}][so](so3)[right of=sp2]{};
                      \node[label=left:{2}][sp](sp2')[above of=so3]{};
                      \node[label=below:{1}][u](u1)[right of=so3]{};
                      \node[label=below:{1}][u](u1')[right of=u1]{};
                      \node[label=above:{2}][uf](uf)[above of=u1]{};
                      \node[label=above:{1}][uf](uf')[above of=u1']{};
                      \node[label=above:{1}][so](sof)[above of=sp2']{};
                      \draw(o1)--(sp2);
                      \draw(sp2)--(so3);
                      \draw(so3)--(sp2');
                      \draw(sp2')--(sof);
                      \draw(so3)--(u1);
                      \draw(u1)--(u1');
                      \draw(u1)--(uf);
                      \draw(u1')--(uf');
             \end{tikzpicture}
         \end{scriptsize}
    \end{array} ~.
    \label{SO8:MQ102}
\end{equation}
The unrefined Coulomb and Higgs branch Hilbert series of \eqref{SO8:MQ210}, \eqref{SO8:MQ201} and \eqref{SO8:MQ102} match and are given in table \ref{CoulombHSOSp} and table \ref{HiggsHSOSp} respectively. We further propose the following highest weight generating function of the Coulomb branch of these quivers:
\begin{equation}
    \HWG_{\mathcal{C}}=\PE\left[\left(1+\mu_1^2+\nu^2\right)t^2+\mu_2^2t^4+\left(q+q^{-1}\right)\mu_2\nu t^6-\mu_2^2\nu^2t^{12}\right]\;,
\end{equation}
where $\mu_i$ are USp$(4)$ highest weight fugacities, $\nu$ is an SU$(2)$ highest weight fugacity, and $q$ is the U(1) charge. 
\subsection{SO(8)+2\textbf{S}+2\textbf{C} \texorpdfstring{$\longleftrightarrow$}{TEXT} SO(8)+2\textbf{S}+2\textbf{V}}
The brane web for SO(8)+2\textbf{S}+2\textbf{C} at infinite coupling is given by
\begin{equation}
    \begin{array}{c}
         \begin{scriptsize}
             \begin{tikzpicture}
                      \draw[thick, dashed](0,0)--(1,0);
                      \draw[thick, dashed](9,0)--(10,0);
                      \draw[thick](1,0)--(9,0);
                      \node[7brane]at(1,0){};
                      \node[7brane]at(2,0){};
                      \node[7brane]at(3,0){};
                      \node[7brane]at(4,0){};
                      \node[7brane]at(6,0){};
                      \node[7brane]at(7,0){};
                      \node[7brane]at(8,0){};
                      \node[7brane]at(9,0){};
                      \draw[thick](5,0)--(6,1);
                      \draw[thick](5,0)--(4,1);
                      \node[label=above:{$2(1,1)$}][7brane]at(6,1){};
                      \node[label=above:{$2(1,-1)$}][7brane]at(4,1){};
                      \node[label=below:{O5$^-$}]at(0.5,0){};
                      \node[label=below:{$\frac{1}{2}$}]at(1.5,0){};
                      \node[label=below:{$\frac{3}{2}$}]at(3.5,0){};
                      \node[label=below:{$1$}]at(2.5,0){};
                      \node[label=below:{$2$}]at(5.5,0){};
                      \node[label=below:{$2$}]at(4.5,0){};
                      \node[label=below:{$\frac{3}{2}$}]at(6.5,0){};
                      \node[label=below:{$1$}]at(7.5,0){};
                      \node[label=below:{$\frac{1}{2}$}]at(8.5,0){};
                      \node[label=below:{O5$^-$}]at(9.5,0){};
             \end{tikzpicture}
         \end{scriptsize}
    \end{array}\;.
\end{equation}
From this, we obtain the following magnetic quiver:
\begin{equation}
\MQ_{1;\, \text{SO}(8)}^{2,2,0} = \begin{array}{c}
     \begin{scriptsize}
         \begin{tikzpicture}
              \node[label=below:{2}][so](so2){};
              \node[label=below:{2}][sp](sp2)[right of=so2]{};
              \node[label=below:{4}][so](so4)[right of=sp2]{};
              \node[label=below:{2}][sp](sp2')[right of=so4]{};
              \node[label=below:{2}][so](so2')[right of=sp2']{};
              \node[label=left:{2}][u](u2)[above of=so4]{};
              \draw(so2)--(sp2);
              \draw(sp2)--(so4);
              \draw(so4)--(sp2');
              \draw(sp2')--(so2');
              \draw(so4)--(u2);
              \draw (u2)to[out=45,in=135,loop,looseness=10](u2);
              \node[label=below:{$2\;\wedge^2$}][above of=u2]{};
         \end{tikzpicture}
     \end{scriptsize}
\end{array} \quad;\quad
\MQ_{2;\, \text{SO}(8)}^{2,2,0} = \begin{array}{c}
     \begin{scriptsize}
         \begin{tikzpicture}
              \node[label=below:{2}][so](so2){};
              \node[label=below:{2}][sp](sp2)[right of=so2]{};
              \node[label=below:{2}][sp](sp2')[right of=sp2]{};
              \node[label=below:{2}][so](so2')[right of=sp2']{};
              \draw(so2)--(sp2);
              \draw[double distance=2pt](sp2)--(sp2');
              \draw(sp2')--(so2');
         \end{tikzpicture}
     \end{scriptsize}
\end{array}
\;.
\label{SO8:MQ220}
\end{equation}
On the other hand, SO(8) triality implies the equivalence of SO(8)+2\textbf{S}+2\textbf{C} with SO(8)+2\textbf{S}+2\textbf{V}. The latter theory has a brane web at infinite coupling given by
\begin{equation}
    \begin{array}{c}
         \begin{scriptsize}
             \begin{tikzpicture}
                  \draw[thick, dashed](2,0)--(3,0);
                      \draw[thick, dashed](7,0)--(8,0);
                      \draw[thick](3,0)--(7,0);
                      \node[7brane]at(3,0){};
                      \node[7brane]at(4,0){};
                      \node[7brane]at(6,0){};
                      \node[7brane]at(7,0){};
                      \node[7brane]at(6,1){};
                      \node[7brane]at(4,1){};
                      \draw[thick](5,0)--(7,2);
                      \draw[thick](5,0)--(3,2);
                      \node[label=above:{$(1,1)$}][7brane]at(7,2){};
                      \node[label=above:{$(1,-1)$}][7brane]at(3,2){};
                      \node[label=below:{O5$^-$}]at(2.5,0){};
                      \node[label=below:{$\frac{3}{2}$}]at(3.5,0){};
                      \node[label=below:{$2$}]at(5.5,0){};
                      \node[label=below:{$2$}]at(4.5,0){};
                      \node[label=below:{$\frac{3}{2}$}]at(6.5,0){};
                      \node[label=below:{O5$^-$}]at(7.5,0){};
                      \node[label=above:{1}]at(6.5,1.5){};
                      \node[label=above:{2}]at(5.5,.5){};
                      \node[label=above:{1}]at(3.5,1.5){};
                      \node[label=above:{2}]at(4.5,.5){};
             \end{tikzpicture}
         \end{scriptsize}
    \end{array}\;,
\end{equation}
From which we obtain the corresponding magnetic quiver to be
\begin{align}
  \MQ_{1;\, \text{SO}(8)}^{2,0,2} &=  \begin{array}{c}
         \begin{scriptsize}
             \begin{tikzpicture}
                      \node[label=below:{2}][sp](sp2){};
                      \node[label=below:{4}][so](so4)[right of=sp2]{};
                      \node[label=below:{2}][sp](sp2')[right of=so4]{};
                      \node[label=below left:{2}][u](u2)[above of=so4]{};
                      \node[label=above:{1}][u](u1)[left of=u2]{};
                      \node[label=above:{1}][u](u1')[right of=u2]{};
                      \draw(sp2)--(so4);
                      \draw(so4)--(sp2');
                      \draw(so4)--(u2);
                      \draw(u2)--(u1);
                      \draw(u2)--(u1');
                      \draw (u2)to[out=45,in=135,loop,looseness=10](u2);
                      \node[label=below:{$2\;\wedge^2$}][above of=u2]{};
             \end{tikzpicture}
         \end{scriptsize}
    \end{array} \quad\contour{black}{$\xrightarrow{\text{B2G}}$}\quad \begin{array}{c}
         \begin{scriptsize}
             \begin{tikzpicture}
                      \node[label=below:{1}][u](sp2){};
                      \node[label=below:{4}][so](so4)[right of=sp2]{};
                      \node[label=below:{1}][u](sp2')[right of=so4]{};
                      \node[label=below left:{2}][u](u2)[above of=so4]{};
                      \node[label=above:{1}][u](u1)[left of=u2]{};
                      \node[label=above:{1}][u](u1')[right of=u2]{};
                      \draw(sp2)--node[below]{$\frac{1}{2}$}++(so4);
                      \draw(so4)--node[below]{$\frac{1}{2}$}++(sp2');
                      \draw(so4)--(u2);
                      \draw(u2)--(u1);
                      \draw(u2)--(u1');
                      \draw (u2)to[out=45,in=135,loop,looseness=10](u2);
                      \node[label=below:{$2\;\wedge^2$}][above of=u2]{};
             \end{tikzpicture}
         \end{scriptsize}
    \end{array} \nonumber \\
       \MQ_{2;\, \text{SO}(8)}^{2,0,2} &= \begin{array}{c}
         \begin{scriptsize}
             \begin{tikzpicture}
                      \node[label=below:{1}][u](so2){};
                      \node[label=below:{2}][sp](sp2)[right of=so2]{};
                      \node[label=below:{2}][sp](sp2')[right of=sp2]{};
                      \node[label=below:{1}][u](u1)[right of=sp2']{};
                      \draw(so2)--(sp2);
                      \draw[double distance=2pt](sp2)--(sp2');
                      \draw(sp2')--(u1);
             \end{tikzpicture}
         \end{scriptsize}
    \end{array}
    \label{SO8:MQ202}
\end{align}
The unrefined Coulomb and Higgs branch Hilbert series of $\MQ_{i;\, \text{SO}(8)}^{2,2,0}$ and $\MQ_{i;\, \text{SO}(8)}^{2,0,2}$ match and are given in table \ref{CoulombHSOSp} and table \ref{HiggsHSOSp} respectively. The HWG's of the Coulomb branches for the second cones take a very simple form and are given as
\begin{equation}
  \HWG_{\mathcal{C}}(\MQ_{2;\, \text{SO}(8)}^{2,2,0}) = \HWG_{\mathcal{C}}(\MQ_{2;\, \text{SO}(8)}^{2,0,2}) =  \PE\left[\mu_1^2t^2+\left(\mu_2+\mu_2^2\right)t^4 +\mu_1^2t^6+\mu_1^2\mu_2t^8-\mu_1^4\mu_2^2t^{16}\right]
\end{equation}
where $\mu_i$ are USp$(4)$ highest weight fugacities.
\subsection{SO(8)+1\textbf{S}+1\textbf{C}+2\textbf{V} \texorpdfstring{$\longleftrightarrow$}{TEXT} SO(8)+2\textbf{S}+1\textbf{C}+1\textbf{V}}
The brane web for the SCFT limit of SO$(8)$+1\textbf{S}+1\textbf{C}+2\textbf{V} is given by
\begin{equation}
    \begin{array}{c}
         \begin{scriptsize}
             \begin{tikzpicture}
                      \draw[thick](0,0)--(-4,0);
                      \draw[thick](0,0)--(4,0);
                      \draw[thick,dashed](-5,0)--(-4,0);
                      \draw[thick,dashed](5,0)--(4,0);
                      \node[7brane]at(-1,0){};
                      \node[7brane]at(-2,0){};
                      \node[7brane]at(-3,0){};
                      \node[7brane]at(-4,0){};
                      \node[7brane]at(1,0){};
                      \node[7brane]at(2,0){};
                      \node[7brane]at(3,0){};
                      \node[7brane]at(4,0){};
                      \node[7brane]at(0,1){};
                      \node[label=above:{$(1,1)$}][7brane]at(1,1){};
                      \draw[thick](0,0)--(1,1);
                      \draw[thick](0,0)--(0,1);
                      \node[label=below:{$\frac{1}{2}$}]at(3.5,0){};
                      \node[label=below:{$1$}]at(2.5,0){};
                      \node[label=below:{$\frac{3}{2}$}]at(1.5,0){};
                      \node[label=below:{$2$}]at(.5,0){};
                      \node[label=below:{$1$}]at(-3.5,0){};
                      \node[label=below:{$1$}]at(-2.5,0){};
                      \node[label=below:{$2$}]at(-1.5,0){};
                      \node[label=below:{$2$}]at(-.5,0){};
                      \node[label=below:{O5$^+$}]at(-4.5,0){};
                      \node[label=below:{O5$^-$}]at(4.5,0){};
                      \node[label=left:{$1$}]at(0,.5){};
                      \node[label=right:{$2$}]at(0.5,.5){};
             \end{tikzpicture}
         \end{scriptsize}
    \end{array}
\end{equation}
The magnetic quivers associated to this brane system are
\begin{align}
   \MQ_{1;\, \text{SO}(8)}^{1,1,2} &= \begin{array}{c}
         \begin{scriptsize}
             \begin{tikzpicture}
                      \node[label=below:{1}][so](o1){};
                      \node[label=below:{2}][sp](sp2)[right of=o1]{};
                      \node[label=below:{3}][so](so3)[right of=sp2]{};
                      \node[label=below:{4}][sp](sp4)[right of=so3]{};
                      \node[label=below:{3}][so](so3')[right of=sp4]{};
                      \node[label=below:{2}][sp](sp2')[right of=so3']{};
                      \node[label=below:{2}][so](so2)[right of=sp2']{};
                      \node[label=left:{1}][u](u1)[above of=sp4]{};
                     \node[label=above:{2}][uf](uf)[above of=u1]{}; \node[label=above:{1}][sof](sof)[above of=sp2']{};
                      \draw(o1)--(sp2);
                      \draw(sp2)--(so3);
                      \draw(so3)--(sp4);
                      \draw(sp4)--(u1);
                      \draw(u1)--(uf);
                      \draw(sp4)--(so3');
                      \draw(so3')--(sp2');
                      \draw(sp2')--(sof);
                      \draw(sp2')--(so2);
             \end{tikzpicture}
         \end{scriptsize}
    \end{array}  \nonumber \\
    \MQ_{2;\, \text{SO}(8)}^{1,1,2} &=
    \begin{array}{c}
         \begin{scriptsize}
             \begin{tikzpicture}
                      \node[label=below:{1}][so](o1){};
                      \node[label=below:{2}][sp](sp2)[right of=o1]{};
                      \node[label=below:{3}][so](so3)[right of=sp2]{};
                      \node[label=below:{2}][sp](sp2')[right of=so3]{};
                      \node[label=left:{2}][sp](sp22)[above of=so3]{};
                      \node[label=below:{1}][so](sof)[right of=sp2']{};
                      \node[label=above:{5}][sof](soff)[above of=sp22]{};
                      \draw(o1)--(sp2);
                      \draw(sp2)--(so3);
                      \draw(so3)--(sp2');
                      \draw(sp2')--(sof);
                      \draw(so3)--(sp22);
                      \draw(sp22)--(soff);
             \end{tikzpicture}
         \end{scriptsize}
    \end{array}
    \label{SO8:MQ112}
\end{align}
The brane web for the SCFT limit of SO$(8)$+2\textbf{S}+1\textbf{C}+1\textbf{V} is given by
\begin{equation}
    \begin{array}{c}
         \begin{scriptsize}
             \begin{tikzpicture}
                      \draw[thick](0,0)--(-2,0);
                      \draw[thick](0,0)--(4,0);
                      \draw[thick](0,0)--(-2,2);
                      \draw[thick](0,0)--(1,1);
                      \draw[thick,dashed](5,0)--(4,0);
                      \draw[thick,dashed](-3,0)--(-2,0);
                      \node[7brane]at(-1,0){};
                      \node[7brane]at(-2,0){};
                      \node[7brane]at(1,0){};
                      \node[7brane]at(2,0){};
                      \node[7brane]at(3,0){};
                      \node[7brane]at(4,0){};
                      \node[7brane]at(-1,1){};
                      \node[label=above:{$(1,-1)$}][7brane]at(-2,2){};
                      \node[label=above:{$(1,1)$}][7brane]at(1,1){};
                      \node[label=below:{O5$^-$}]at(-2.5,0){};
                      \node[label=below:{$\frac{3}{2}$}]at(-1.5,0){};
                      \node[label=below:{$2$}]at(-.5,0){};
                      \node[label=below:{$\frac{1}{2}$}]at(3.5,0){};
                      \node[label=below:{$1$}]at(2.5,0){};
                      \node[label=below:{$\frac{3}{2}$}]at(1.5,0){};
                      \node[label=below:{$2$}]at(.5,0){};
                      \node[label=above:{1}]at(-1.5,1.5){};
                      \node[label=above:{2}]at(-.5,.5){};
                      \node[label=above:{2}]at(.5,.5){};
                      \node[label=below:{O5$^-$}]at(4.5,0){};
             \end{tikzpicture}
         \end{scriptsize}
    \end{array}
\end{equation}
The magnetic quiver that we read off here is 
\begin{align}
  \MQ_{1;\, \text{SO}(8)}^{2,1,1} &=  \begin{array}{c}
         \begin{scriptsize}
             \begin{tikzpicture}
                      \node[label=below:{2}][sp](sp2){};
                      \node[label=below:{4}][so](so4)[right of=sp2]{};
                      \node[label=below:{2}][sp](sp2')[right of=so4]{};
                      \node[label=below:{2}][so](so2)[right of=sp2']{};
                      \node[label=right:{2}][u](u2)[above of=so4]{};
                      \node[label=right:{1}][u](u1)[above of=u2]{};
                      \draw(u2)to[out=135,in=-135,loop,looseness=10](u2);
                      \node[left of=u2]{2 $\wedge^2$};
                      \draw(u1)--(u2);
                      \draw(u2)--(so4);
                      \draw(sp2)--(so4);
                      \draw(so4)--(sp2');
                      \draw(sp2')--(so2);
             \end{tikzpicture}
         \end{scriptsize}
    \end{array} \quad\contour{black}{$\xrightarrow{\text{B2G}}$}\quad \begin{array}{c}
         \begin{scriptsize}
             \begin{tikzpicture}
                      \node[label=below:{1}][u](sp2){};
                      \node[label=below:{4}][so](so4)[right of=sp2]{};
                      \node[label=below:{2}][sp](sp2')[right of=so4]{};
                      \node[label=below:{2}][so](so2)[right of=sp2']{};
                      \node[label=right:{2}][u](u2)[above of=so4]{};
                      \node[label=right:{1}][u](u1)[above of=u2]{};
                      \draw(u2)to[out=135,in=-135,loop,looseness=10](u2);
                      \node[left of=u2]{2 $\wedge^2$};
                      \draw(u1)--(u2);
                      \draw(u2)--(so4);
                      \draw(sp2)--node[below]{$\frac{1}{2}$}++(so4);
                      \draw(so4)--(sp2');
                      \draw(sp2')--(so2);
             \end{tikzpicture}
         \end{scriptsize}
    \end{array} \nonumber \\
    \MQ_{2;\, \text{SO}(8)}^{2,1,1} &= \begin{array}{c}
         \begin{scriptsize}
             \begin{tikzpicture}
                      \node[label=below:{2}][so](so2){};
                      \node[label=below:{2}][sp](sp2)[right of=so2]{};
                      \node[label=below:{2}][sp](sp2')[right of=sp2]{};
                      \node[label=above:{1}][u](u1)[above of=sp2]{};
                      \draw(so2)--(sp2);
                      \draw[double distance=2pt](sp2)--(sp2');
                      \draw(sp2)--(u1);
             \end{tikzpicture}
         \end{scriptsize}
    \end{array}
    \label{SO8:MQ211}
\end{align}
The unrefined Coulomb and Higgs branch Hilbert series of $\MQ_{i;\, \text{SO}(8)}^{1,1,2}$ and $\MQ_{i;\, \text{SO}(8)}^{2,1,1}$ match and are given in table \ref{CoulombHSOSp} and table \ref{HiggsHSOSp} respectively. The HWG's for the Coulomb branches for the second cones takes a very simple form
\begin{equation}
   \HWG_{\mathcal{C}}(\MQ_{2;\, \text{SO}(8)}^{1,1,2}) = \HWG_{\mathcal{C}}(\MQ_{2;\, \text{SO}(8)}^{2,1,1}) = \PE\left[\mu_1^2t^2+\left(1+\mu_2+\mu_2^2\right)t^4 +\mu_2t^6-\mu_2^2t^{12}\right]\;,
\end{equation}
where the $\mu_i$ are USp(4) highest weight fugacities.
\subsection{SO(8)+2\textbf{S}+2\textbf{C}+1\textbf{V} \texorpdfstring{$\longleftrightarrow$}{TEXT} SO(8)+2\textbf{S}+1\textbf{C}+2\textbf{V}}
The brane web for the SCFT limit of SO$(8)$+2\textbf{S}+2\textbf{C}+1\textbf{V} takes the form
\begin{equation}
    \begin{array}{c}
         \begin{scriptsize}
             \begin{tikzpicture}
                      \draw[thick](6,0)--(-4,0);
                      \draw[thick,dashed](-5,0)--(-4,0);
                      \draw[thick,dashed](6,0)--(7,0);
                      \node[7brane]at(-4,0){};
                      \node[7brane]at(-3,0){};
                      \node[7brane]at(-2,0){};
                      \node[7brane]at(-1,0){};
                      \node[7brane]at(6,0){};
                      \node[7brane]at(5,0){};
                      \node[7brane]at(4,0){};
                      \node[7brane]at(3,0){};
                      \node[7brane]at(2,0){};
                      \node[7brane]at(1,0){};
                      \node[7brane]at(0,1){};
                      \node[label=above:{$(1,-1)$}][7brane]at(-1,1){};
                      \draw[thick](0,0)--(0,1);
                      \draw[thick](0,0)--(-1,1);
                      \node[label=below:{$\frac{1}{2}$}]at(-3.5,0){};
                      \node[label=below:{$1$}]at(-2.5,0){};
                      \node[label=below:{$\frac{3}{2}$}]at(-1.5,0){};
                      \node[label=below:{$2$}]at(-.5,0){};
                      \node[label=below:{$4$}]at(.5,0){};
                      \node[label=below:{$\frac{7}{2}$}]at(1.5,0){};
                      \node[label=below:{$2$}]at(2.5,0){};
                      \node[label=below:{$\frac{3}{2}$}]at(3.5,0){};
                      \node[label=below:{$1$}]at(4.5,0){};
                      \node[label=below:{$\frac{1}{2}$}]at(5.5,0){};
                      \node[label=below:{O5$^-$}]at(6.5,0){};
                      \node[label=below:{O5$^-$}]at(-4.5,0){};
                      \node[label=right:{$2$}]at(0,0.5){};
                      \node[label=left:{$2$}]at(-.5,0.5){};
             \end{tikzpicture}
         \end{scriptsize}
    \end{array}
\end{equation}
The magnetic quivers associated to this brane system are
\begin{align}
  \MQ_{1;\, \text{SO}(8)}^{2,2,1} &= \begin{array}{c}
         \begin{scriptsize}
             \begin{tikzpicture}
                      \node[label=below:{2}][so](so2){};
                      \node[label=below:{2}][sp](sp2)[right of=so2]{};
                      \node[label=below:{4}][so](so4)[right of=sp2]{};
                      \node[label=below:{6}][sp](sp6)[right of=so4]{};
                      \node[label=below:{4}][so](so4')[right of=sp6]{};
                      \node[label=below:{2}][sp](sp2')[right of=so4']{};
                      \node[label=below:{2}][so](so2')[right of=sp2']{};
                      \node[label=left:{2}][u](u2)[above of=sp6]{};
                      \draw(so2)--(sp2);
                      \draw(sp2)--(so4);
                      \draw(so4)--(sp6);
                      \draw(sp6)--(so4');
                      \draw(so4')--(sp2');
                      \draw(sp2')--(so2');
                      \draw(sp6)--(u2);
                      \draw(u2)to[out=45,in=135,loop,looseness=10](u2);
                      \node[label=below:{1 $\wedge^2$}][above of=u2]{};
             \end{tikzpicture}
         \end{scriptsize}
    \end{array} \nonumber \\
   \MQ_{2;\, \text{SO}(8)}^{2,2,1} &= \begin{array}{c}
         \begin{scriptsize}
             \begin{tikzpicture}
                      \node[label=below:{2}][so](so2){};
                      \node[label=below:{2}][sp](sp2)[right of=so2]{};
                      \node[label=below:{4}][so](so4)[right of=sp2]{};
                      \node[label=below:{4}][sp](sp4)[right of=so4]{};
                      \node[label=below:{4}][so](so4')[right of=sp4]{};
                      \node[label=above:{2}][sp](sp22)[above of=so4]{};
                      \node[label=above:{2}][sp](sp22')[above of=so4']{};
                      \node[label=below:{2}][sp](sp2')[right of=so4']{};
                      \node[label=below:{2}][so](so2')[right of=sp2']{};
                      \draw(so2)--(sp2);
                      \draw(sp2)--(so4);
                      \draw(so4)--(sp22);
                      \draw(sp22)--(sp22');
                      \draw(sp22')--(so4');
                      \draw(so4)--(sp4);
                      \draw(so4')--(sp4);
                      \draw(so4')--(sp2');
                      \draw(sp2')--(so2');
             \end{tikzpicture}
         \end{scriptsize}
    \end{array}
    \label{SO8:MQ221}
\end{align}
If, instead, we take the SCFT limit of the brane web for SO$(8)$+2\textbf{S}+1\textbf{C}+2\textbf{V}, given by
\begin{equation}
    \begin{array}{c}
         \begin{scriptsize}
             \begin{tikzpicture}
                      \draw[thick](-2,0)--(6,0);
                      \draw[thick,dashed](-2,0)--(-3,0);
                      \draw[thick,dashed](6,0)--(7,0);
                      \draw[thick](0,0)--(-2,2);
                      \draw[thick](0,0)--(0,1);
                      \node[label=above:{$(1,-1)$}][7brane]at(-2,2){};
                      \node[7brane]at(-1,1){};
                      \node[7brane]at(0,1){};
                      \node[7brane]at(-2,0){};
                      \node[7brane]at(-1,0){};
                      \node[7brane]at(1,0){};
                      \node[7brane]at(2,0){};
                      \node[7brane]at(3,0){};
                      \node[7brane]at(4,0){};
                      \node[7brane]at(5,0){};
                      \node[7brane]at(6,0){};
                      \node[label=below:{$\frac{3}{2}$}]at(-1.5,0){};
                      \node[label=below:{$2$}]at(-.5,0){};
                      \node[label=below:{$4$}]at(.5,0){};
                      \node[label=below:{$\frac{7}{2}$}]at(1.5,0){};
                      \node[label=below:{$2$}]at(2.5,0){};
                      \node[label=below:{$\frac{3}{2}$}]at(3.5,0){};
                      \node[label=below:{$1$}]at(4.5,0){};
                      \node[label=below:{$\frac{1}{2}$}]at(5.5,0){};
                      \node[label=below:{O5$^-$}]at(6.5,0){};
                      \node[label=below:{O5$^-$}]at(-2.5,0){};
                      \node[label=left:{$1$}]at(-1.5,1.5){};
                      \node[label=left:{$2$}]at(-.5,.5){};
                      \node[label=right:{$2$}]at(0,.5){};
             \end{tikzpicture}
         \end{scriptsize}
    \end{array}\;,
\end{equation}
we obtain the following alternative magnetic quivers
\begin{align}
 \MQ_{1;\, \text{SO}(8)}^{2,1,2} &= \begin{array}{c}
         \begin{scriptsize}
             \begin{tikzpicture}
                      \node[label=below:{2}][sp](sp2){};
                      \node[label=below:{4}][so](so4)[right of=sp2]{};
                      \node[label=below:{6}][sp](sp6)[right of=so4]{};
                      \node[label=below:{4}][so](so4')[right of=sp6]{};
                      \node[label=below:{2}][sp](sp2')[right of=so4']{};
                      \node[label=below:{2}][so](so2')[right of=sp2']{};
                      \node[label=left:{2}][u](u2)[above of=sp6]{};
                      \node[label=left:{1}][u](u1)[above of=u2]{};
                      \draw(so2)--(sp2);
                      \draw(sp2)--(so4);
                      \draw(so4)--(sp6);
                      \draw(sp6)--(so4');
                      \draw(so4')--(sp2');
                      \draw(sp2')--(so2');
                      \draw(sp6)--(u2);
                      \draw(u2)--(u1);
                      \draw(u2)to[out=45,in=-45,loop,looseness=10](u2);
                      \node[right of=u2]{1 $\wedge^2$};
             \end{tikzpicture}
         \end{scriptsize}
    \end{array} \nonumber \\
  \MQ_{2;\, \text{SO}(8)}^{2,1,2} &= \begin{array}{c}
         \begin{scriptsize}
             \begin{tikzpicture}
                      \node[label=below:{2}][sp](sp2){};
                      \node[label=below:{4}][so](so4)[right of=sp2]{};
                      \node[label=below:{4}][sp](sp4)[right of=so4]{};
                      \node[label=below:{4}][so](so4')[right of=sp4]{};
                      \node[label=above:{2}][sp](sp22)[above of=so4]{};
                      \node[label=above:{2}][sp](sp22')[above of=so4']{};
                      \node[label=below:{2}][sp](sp2')[right of=so4']{};
                      \node[label=below:{2}][so](so2')[right of=sp2']{};
                      \node[label=above:{1}][u](u1)[right of=sp22']{};
                      \draw(u1)--(sp22');
                      \draw(sp2)--(so4);
                      \draw(so4)--(sp22);
                      \draw(sp22)--(sp22');
                      \draw(sp22')--(so4');
                      \draw(so4)--(sp4);
                      \draw(so4')--(sp4);
                      \draw(so4')--(sp2');
                      \draw(sp2')--(so2');
             \end{tikzpicture}
         \end{scriptsize}
    \end{array}
    \label{SO8:MQ212}
\end{align}
The unrefined Coulomb and Higgs branch Hilbert series of $\MQ_{i;\, \text{SO}(8)}^{2,2,1}$ and $\MQ_{i;\, \text{SO}(8)}^{2,1,2}$ match and are given in table \ref{CoulombHSOSp} and table \ref{HiggsHSOSp} respectively. 

\section{Conclusion}\label{conclusions}
In this paper, we studied the Higgs branches of 5d SCFTs, that admit deformations to SO(6) or SO(8) gauge theories with matter in either the vector, spinor, or conjugate spinor representation of the gauge group. We used the brane configurations engineering these theories, to find the corresponding magnetic quivers. The magnetic quivers for SO(6) theories were verified by comparison with those of SU(4) theories, while we used SO(8) triality, to provide consistency checks of the magnetic quivers proposed for the SO(8) theories. The agreement of the different magnetic quivers for a given SCFT, were shown to hold at the level of the unrefined Hilbert series, for all cases, on both the Coulomb and the Higgs branch of the moduli space. The unrefined Hilbert series for both the Coulomb and the Higgs branch Hilbert series of all magnetic quivers studied in this work are collected in appendix \ref{appendixA}. Some magnetic quivers encountered in this work, involved bad symplectic gauge nodes. We were able to assign a Hilbert series for both Coulomb and Higgs branches of these theories, developing a new technique which is referred to as B2G in the main text, which uses a local mirror description of the effective theory around the most singular locus of the moduli space. Details of these computational techniques will be the content of a future publication \cite{HSbad}. Since the Coulomb branch of a bad 3d $\mathcal{N}=4$ USp($2N$) theory is comprised of multiple singular loci, one may wonder whether a 5d Higgs branch whose magnetic quiver contains bad USp($2N$) nodes, has a similar structure on its Higgs branch. However the output of our study seems to suggest otherwise. In particular, we found that for the theories under our construction, the local geometry near one of the two singular points on the Coulomb branch of the magnetic quiver, can capture the full Higgs branch of the 5d theory. This conclusion follows because for every magnetic quiver with a bad node, we have a dual magnetic quiver without any bad nodes. It would be interesting to find further checks of this claim.

We encountered brane webs whose magnetic quivers involve second rank symmetric and second rank antisymmetric tensors. We provided a formula for computing the number of hypermultiplets in the (anti)symmetric representations, which generalises previous rules for extracting magnetic quivers from brane webs with O5-planes in \cite{Bourget:2020gzi,Akhond:2020vhc}. Despite these improvements, there is still no completely systematic algorithm to extract magnetic quivers for 5d SCFTs engineered using O5-planes. Perhaps the most urgent question to address, is the magnetic quivers for SO($2r+1$) theories, which are engineered using $\widetilde{\text{O5}}$-planes. One might be able to use the results of \cite{Akhond:2021ffo} as a guide for such a study. Another subtle issue, that will eventually need to be addressed, is developing a more systematic method for distinguishing different theories whose 5-brane web at the SCFT limit is identical.

\acknowledgments
We thank Guillermo Arias-Tamargo, Stefano Cremonesi Julius Grimminger, Amihay Hanany, Alessandro Mininno and Zhenghao Zhong for interesting discussions. SK acknowledges the hospitality at APCTP where part of this work was done. MA is supported by STFC grant ST/S505778/1. F.C. is supported by STFC consolidated grant ST/T000708/1. SD is supported by the NSFC grants No. 12050410249 and No. 11975158. The work of HH is supported in part by JSPS KAKENHI Grant Number JP18K13543. SK and FY are partially supported by the Fundamental Research Funds for the Central Universities 2682021ZTPY043. FY is supported by the NSFC grant No. 11950410490 and by Start-up research grant YH1199911312101.
\vspace{1cm}
\appendix
\section{Unrefined Hilbert series results for OSp quivers}\label{appendixA}

\captionsetup{width=15cm}
\begin{longtable}{|c|C{6.11cm}|C{6.10cm}|}
\caption{Unrefined Coulomb branch Hilbert series for the orthosymplectic quivers.}
			\label{CoulombHSOSp} \\ \hline
    & \multicolumn{2}{c|}{$\text{HS}_{\mathcal{C}}(t)=\text{HS}_{\mathcal{C}}(t;\vec{m} \in \mathbb{Z})+\text{HS}_{\mathcal{C}}(t;\vec{m} \in \mathbb{Z}+\tfrac{1}{2})$} \\ \cline{2-3}
  \multirow{-2}{*}{Quiver} & $\text{HS}_{\mathcal{C}}(t;\vec{m} \in \mathbb{Z})$ & $\text{HS}_{\mathcal{C}}(t;\vec{m} \in \mathbb{Z}+\tfrac{1}{2})$  \\[0.1cm] \hline
	\hyperref[SO6:MQ110]{$\MQ_1^{1,1,0}$} & \CBHS{1 + 3 t^4 + 2 t^6 + 3 t^8 + t^{12}}{(1-t^2)^2\,(1-t^4) \,(1-t^8)}{1 + 2 t^2 + 7 t^4 + 14 t^6 + 29 t^8 + 46 t^{10}} & \CBHS{2 t^2 + 2 t^4 + 2 t^6 + 2 t^8 + 2 t^{10}}{(1-t^2)^2\,(1-t^4) \,(1-t^8)}{2 t^2 + 6 t^4 + 14 t^6 + 26 t^8 + 46 t^{10}}  \\ \hline
	\hyperref[SO6:MQ200]{$\MQ_1^{2,0,0}$} & \CBHS{1 + t^4}{(1 - t^2)\,(1 - t^4)}{1 + t^2 + 3 t^4 + 3 t^6 + 5 t^8 + 5 t^{10}} & \footnotesize{not required}  \\ \hline
	\hyperref[SO6:MQ200]{$\MQ_2^{2,0,0}$} & \CBHS{1 + t^4}{(1 - t^2)\,(1 - t^4)}{1 + t^2 + 3 t^4 + 3 t^6 + 5 t^8 + 5 t^{10}} & \CBHS{2t^2}{(1 - t^2)\,(1 - t^4)}{2 t^2 + 2 t^4 + 4 t^6 + 4 t^8 + 6 t^{10}}  \\ \hline
	\hyperref[SO6:MQ210]{$\MQ_1^{2,1,0}$}  & \CBHS{P_1(t)}{(1-t^2)^3\,(1-t^4)^2 \,(1-t^8)}{1 + 5 t^2 + 24 t^4 + 72 t^6 + 189 t^8 + 413 t^{10}} & \CBHS{4 t^2 + 6 t^4 + 14 t^6 + 14 t^8 + 14 t^{10} + 6 t^{12} + 4 t^{14}}{(1-t^2)^3\,(1-t^4)^2 \,(1-t^8)}{4 t^2 + 18 t^4 + 64 t^6 + 168 t^8 + 388 t^{10}}  \\ \hline
	\hyperref[SO6:MQ220]{$\MQ_1^{2,2,0}$}  & \CBHS{P_2(t)}{(1-t^2)^5\,(1-t^4)^4 \,(1-t^8)}{1 + 8 t^2 + 72 t^4 + 371 t^6 + 1598 t^8 + 5510 t^{10}} & \CBHS{P_3(t)}{(1-t^2)^5\,(1-t^4)^4 \,(1-t^8)}{8 t^2 + 60 t^4 + 364 t^6 + 1536 t^8 + 5464 t^{10}}  \\ \hline
	\hyperref[SO6:MQ220]{$\MQ_2^{2,2,0}$}  & \CBHS{1+t^4}{(1-t^2)\,(1-t^4)}{1 + t^2 + 3 t^4 + 3 t^6 + 5 t^8 + 5 t^{10}} & \footnotesize{not required}  \\ \hline
\hyperref[SO6:MQtilde220]{$\widehat{\MQ}_1^{2,2,0}$}  & \CBHS{P_4(t)}{(1-t^2)^5\,(1-t^4)^5}{1 + 16 t^2 + 132 t^4 + 735 t^6 + 3134 t^8 \\+ 10974 t^{10}} & \footnotesize{not required}  \\ \hline
 \hyperref[SO6:MQtilde220]{$\widehat{\MQ}_2^{2,2,0}$}  & \CBHS{1+t^4}{(1-t^2)\,(1-t^4)}{1 + t^2 + 3 t^4 + 3 t^6 + 5 t^8 + 5 t^{10}} & \footnotesize{not required}  \\\hline
\hyperref[SO6:MQ320]{$\MQ_1^{3,2,0}$}  & \footnotesize{$1 + 17 t^2 + 184 t^4 + 1446 t^6 + 8758 t^8 + 43000 t^{10} + 
 178362 t^{12} + 644654 t^{14} + 2079047 t^{16} + 6092795 t^{18} + 
 16458838 t^{20}+\ldots$} & \footnotesize{$8 t^2 + 136 t^4 + 1248 t^6 + 8072 t^8 + 40952 t^{10} + 172888 t^{12} + 
 631376 t^{14} + 2049176 t^{16} + 6030000 t^{18} + 16333832 t^{20}+\ldots$}  \\\hline
\hyperref[SO6:MQ320]{$\MQ_2^{3,2,0}$}  & \CBHS{P_5(t)}{(1-t^2)^5\,(1-t^4)\,(1-t^8)^4}{1 + 17 t^2 + 139 t^4 + 751 t^6 + 3148 t^8\\ + 10894 t^{10}}  & \CBHS{P_6(t)}{(1-t^2)^5\,(1-t^4)\,(1-t^8)^4}{8 t^2 + 96 t^4 + 624 t^6 + 2832 t^8 + 10232 t^{10}}  \\ \hline
 \hyperref[SO6:MQ330]{$\MQ_1^{3,3,0}$}  & \footnotesize{$1 + 20 t^2 + 350 t^4 + 4199 t^6 + 38358 t^8 + 278738 t^{10} + 
 1683601 t^{12} + 8713628 t^{14} + 39600362 t^{16} + 161030946 t^{18} + 
 594866176 t^{20}+\ldots$} & \footnotesize{$16 t^2 + 320 t^4 + 4096 t^6 + 37920 t^8 + 277504 t^{10} +  1679744 t^{12} + 8704320 t^{14} + 39576496 t^{16} + 160979760 t^{18} + 
 594751936 t^{20}+\ldots$}  \\\hline
 \hyperref[SO6:MQ330]{$\MQ_2^{3,3,0}$}  & \footnotesize{$1 + 20 t^2 + 340 t^4 + 3926 t^6 + 33526 t^8 + 224534 t^{10} + 
 1240034 t^{12} + 5850463 t^{14} + 24223718 t^{16} + 89832720 t^{18} + 
 303191840 t^{20}+\ldots$} & \footnotesize{$16 t^2 + 320 t^4 + 3872 t^6 + 33344 t^8 + 224128 t^{10} + 
 1238960 t^{12} + 5848352 t^{14} + 24218944 t^{16} + 89824160 t^{18} + 
 303174592 t^{20}+\ldots$}  \\ \hline
 \hyperref[SO6:MQ440]{$\MQ_1^{4,4,0}$}  & \footnotesize{$1 + 37 t^2 + 1350 t^4 + 34389 t^6 + 668310 t^8 + 10281564 t^{10} + 
 129992857 t^{12} + 1388266357 t^{14} + 12803207039 t^{16} + 
 103789879656 t^{18} + 750444248396 t^{20}+\ldots$} & \footnotesize{$32 t^2 + 1280 t^4 + 34080 t^6 + 666016 t^8 + 10272864 t^{10} + 
 129948064 t^{12} + 1388117408 t^{14} + 12802606720 t^{16} + 
 103788095648 t^{18} + 750438227104 t^{20}+\ldots$}  \\ \hline
 \hyperref[SO6:MQ201]{$\MQ_1^{2,0,1}$}  & \CBHS{P_7(t)}{(1-t^2)^2\,(1-t^4)\,(1-t^6)^3}{1 + 5 t^2 + 18 t^4 + 58 t^6 + 149 t^8 + 325 t^{10}}  & \CBHS{2t^3 P_8(t)}{(1-t^2)\,(1-t^4)^2\,(1-t^6)^3}{6 t^3 + 26 t^5 + 78 t^7 + 198 t^9}  \\ \hline
 \hyperref[SO6:MQ201]{$\MQ_2^{2,0,1}$}  & \CBHS{1+t^4}{(1-t^2)^4}{1 + 4 t^2 + 11 t^4 + 24 t^6 + 45 t^8 + 76 t^{10}}  & \CBHS{2t^2}{(1-t^2)^4}{2 t^2 + 8 t^4 + 20 t^6 + 40 t^8 + 70 t^{10}}  \\ \hline
 \hyperref[SO6:MQ201]{$\MQ_3^{2,0,1}$}  & \CBHS{1 - 2 t + 3 t^2 - 4 t^3 + 6 t^4 - 4 t^5 + 3 t^6 - 2 t^7 + t^8}{(1-t)^2\,(1-t^4)\,(1-t^6)}{1 + 2 t^2 + 5 t^4 + 4 t^5 + 10 t^6 + 8 t^7 + 17 t^8\\ + 16 t^9 + 28 t^{10}}  & \CBHS{2 t^2 - 2 t^3 + 2 t^4 - 2 t^5 + 2 t^6}{(1-t)^2\,(1-t^4)\,(1-t^6)}{2 t^2 + 2 t^3 + 4 t^4 + 4 t^5 + 8 t^6 + 10 t^7 + 16 t^8\\ + 18 t^9 + 
 26 t^{10}}  \\ \hline
  \hyperref[SO6:MQ111]{$\MQ_1^{1,1,1}$}  & \CBHS{P_9(t)}{(1-t^2)^3\,(1-t^4)\,(1-t^8)^2}{1 + 5 t^2 + 20 t^4 + 60 t^6 + 157 t^8 + 345 t^{10}}  & \CBHS{(1-t^4)P_{10}(t)}{(1-t^2)^5\,(1-t^8)^2}{2 t^2 + 14 t^4 + 50 t^6 + 136 t^8 + 314 t^{10}}  \\ \hline
    \hyperref[SO6:MQ111]{$\MQ_2^{1,1,1}$}  & \CBHS{1 + 2 t^2 + 2 t^4 + 6 t^6 + 2 t^8 + 2 t^{10} + t^{12}}{(1-t^2)^2\,(1-t^6)^2}{1 + 4 t^2 + 9 t^4 + 22 t^6 + 41 t^8 + 66 t^{10}}  & \CBHS{4 t^3 + 4 t^5 + 4 t^7 + 4 t^9}{(1-t^2)^2\,(1-t^6)^2}{4 t^3 + 12 t^5 + 24 t^7 + 48 t^9}  \\ \hline
 \hyperref[SO6:MQ211]{$\MQ_1^{2,1,1}$}  & \footnotesize{$1 + 9 t^2 + 4 t^3 + 57 t^4 + 52 t^5 + 291 t^6 + 312 t^7 + 1172 t^8 + 
 1360 t^9 + 3932 t^{10}+\ldots$} & \footnotesize{$4 t^2 + 10 t^3 + 36 t^4 + 82 t^5 + 208 t^6 + 426 t^7 + 920 t^8 + 
 1708 t^9 + 3276 t^{10}+\ldots$}  \\ \hline
 \hyperref[SO6:MQ211]{$\MQ_2^{2,1,1}$}  & \CBHS{1 + t^2 + 7 t^4 + 6 t^6 + 7 t^8 + t^{10} + t^{12}}{(1-t^2)^3\,(1-t^4)^3}{1 + 4 t^2 + 19 t^4 + 55 t^6 + 146 t^8 + 317 t^{10}} & \CBHS{4 t^2 + 4 t^4 + 8 t^6 + 4 t^8 + 4 t^{10}}{(1-t^2)^3\,(1-t^4)^3}{4 t^2 + 16 t^4 + 56 t^6 + 140 t^8 + 320 t^{10}}  \\ \hline
 \hyperref[SO6:MQ221]{$\MQ_1^{2,2,1}$}  & \footnotesize{$1 + 20 t^2 + 28 t^3 + 220 t^4 + 456 t^5 + 1905 t^6 + 4132 t^7 + 
 13026 t^8 + 27600 t^9 + 72438 t^{10}+\ldots$} & \footnotesize{not required}  \\ \hline
 \hyperref[SO6:MQ221]{$\MQ_2^{2,2,1}$}  & \footnotesize{$1 + 19 t^2 + 24 t^3 + 188 t^4 + 368 t^5 + 1396 t^6 + 2968 t^7 + 
 8302 t^8 + 17168 t^9 + 40474 t^{10} + 79648 t^{11} + 167230 t^{12} + 
 312656 t^{13} + 603338 t^{14} + 1074896 t^{15} + 1943919 t^{16} + 
 3316912 t^{17} + 5693149 t^{18} + 9351600 t^{19} + 15372660 t^{20}+\ldots$} & \footnotesize{not required}  \\ \hline
 \hyperref[SO6:MQhat221]{$\widehat{\MQ}_1^{2,2,1}$}  & \footnotesize{$1 + 12 t^2 + 8 t^3 + 124 t^4 + 200 t^5 + 1033 t^6 + 1912 t^7 + 
 6834 t^8 + 13192 t^9 + 37406 t^{10}+\ldots$} & \footnotesize{$8 t^2 + 20 t^3 + 96 t^4 + 256 t^5 + 872 t^6 + 2220 t^7 + 6192 t^8 + 
 14408 t^9 + 35032 t^{10}+\ldots$}  \\ \hline
 \hyperref[SO6:MQhat221]{$\widehat{\MQ}_2^{2,2,1}$}  & \footnotesize{$1 + 11 t^2 + 8 t^3 + 100 t^4 + 176 t^5 + 724 t^6 + 1448 t^7 + 
 4206 t^8 + 8496 t^9 + 20394 t^{10}+\ldots$} & \footnotesize{$8 t^2 + 16 t^3 + 88 t^4 + 192 t^5 + 672 t^6 + 1520 t^7 + 4096 t^8 + 
 8672 t^9 + 20080 t^{10}+\ldots$}  \\ \hline
 \hyperref[SO6:MQ321]{$\MQ_1^{3,2,1}$}  & \footnotesize{$1 + 21 t^2 + 28 t^3 + 292 t^4 + 710 t^5 + 3576 t^6 + 10234 t^7 + 
 37720 t^8 + 107242 t^9 + 331772 t^{10}+\ldots$} & \footnotesize{$8 t^2 + 24 t^3 + 192 t^4 + 688 t^5 + 2992 t^6 + 10144 t^7 + 
 34904 t^8 + 106960 t^9 + 320096 t^{10}+\ldots$}  \\ \hline
 \hyperref[SO6:MQ331]{$\MQ_1^{3,3,1}$}  & \footnotesize{$1 + 24 t^2 + 36 t^3 + 497 t^4 + 1544 t^5 + 9096 t^6 + 32876 t^7 + 
 141780 t^8 + 501144 t^9 + 1831783 t^{10}+\ldots$} & \footnotesize{$16 t^2 + 48 t^3 + 432 t^4 + 1648 t^5 + 8656 t^6 + 33680 t^7 + 
 139248 t^8 + 505904 t^9 + 1818896 t^{10}+\ldots$}  \\ \hline
 \hyperref[SO6:MQ441]{$\MQ_1^{4,4,1}$}  & \footnotesize{$1 + 46 t^2 + 96 t^3 + 1836 t^4 + 8256 t^5 + 66981 t^6+\ldots$} & \footnotesize{$32 t^2 + 128 t^3 + 1664 t^4 + 8704 t^5 + 64960 t^6+\ldots$}  \\ \hline
 \hyperref[SO6:MQ112]{$\MQ_1^{1,1,2}$}  & \footnotesize{$1 + 18 t^2 + 205 t^4 + 1591 t^6 + 9499 t^8 + 45959 t^{10} + 
 188535 t^{12} + 675381 t^{14} + 2163400 t^{16} + 6305809 t^{18} + 
 16961842 t^{20}+\ldots$} & \footnotesize{not required}  \\ \hline
 \hyperref[SO6:MQ112]{$\MQ_2^{1,1,2}$}  & \footnotesize{$ 1 + 24 t^2 + 8 t^3 + 255 t^4 + 144 t^5 + 1716 t^6 + 1256 t^7 + 
 8594 t^8 + 7312 t^9 + 34872 t^{10} + 32640 t^{11} + 120628 t^{12} + 
 120336 t^{13} + 367968 t^{14} + 383440 t^{15} + 1013621 t^{16} + 
 1088720 t^{17} + 2565512 t^{18} + 2814744 t^{19} + 6045369 t^{20}+\ldots$} & \footnotesize{not required}  \\ \hline
 \hyperref[SO6:MQ112]{$\MQ_3^{1,1,2}$}  & \footnotesize{$1 + 11 t^2 + 8 t^3 + 65 t^4 + 80 t^5 + 295 t^6 + 432 t^7 + 1122 t^8 + 
 1720 t^9 + 3666 t^{10} + 5640 t^{11} + 10564 t^{12} + 16024 t^{13} + 
 27460 t^{14} + 40752 t^{15} + 65445 t^{16} + 94888 t^{17} + 144935 t^{18} + 
 205440 t^{19} + 301571 t^{20}+\ldots$} & \footnotesize{not required}  \\ \hline
 \hyperref[SO6:MQtilde112]{$\widehat{\MQ}_1^{1,1,2}$}  & \footnotesize{$1 + 14 t^2 + 129 t^4 + 895 t^6 + 5071 t^8 + 23887 t^{10} + 96579 t^{12} + 
 343049 t^{14} + 1093344 t^{16} + 3176685 t^{18} + 8527154 t^{20}+\ldots$} & \footnotesize{$4 t^2 + 76 t^4 + 696 t^6 + 4428 t^8 + 22072 t^{10} + 91956 t^{12} +
 332332 t^{14} + 1070056 t^{16} + 3129124 t^{18} + 8434688 t^{20}+\ldots$}  \\ \hline
 \hyperref[SO6:MQtilde112]{$\widehat{\MQ}_2^{1,1,2}$}  & \footnotesize{$1 + 20 t^2 + 179 t^4 + 32 t^5 + 1092 t^6 + 416 t^7 + 5142 t^8 + 
 2816 t^9 + 20024 t^{10} + 13600 t^{11} + 67328 t^{12} + 52544 t^{13} + 
 201208 t^{14} + 172576 t^{15} + 545849 t^{16} + 500352 t^{17} + 
 1365524 t^{18} + 1313248 t^{19} + 3188473 t^{20}+\ldots$} & \footnotesize{$4 t^2 + 8 t^3 + 76 t^4 + 112 t^5 + 624 t^6 + 840 t^7 + 3452 t^8 + 
 4496 t^9 + 14848 t^{10} + 19040 t^{11} + 53300 t^{12} + 67792 t^{13} + 
 166760 t^{14} + 210864 t^{15} + 467772 t^{16} + 588368 t^{17} + 
 1199988 t^{18} + 1501496 t^{19} + 2856896 t^{20}+\ldots$}  \\ \hline
 \hyperref[SO6:MQtilde112]{$\widehat{\MQ}_3^{1,1,2}$}  & \footnotesize{$1 + 11 t^2 + 65 t^4 + 295 t^6 + 1122 t^8 + 3666 t^{10} + 10564 t^{12} + 
 27460 t^{14} + 65445 t^{16} + 144935 t^{18} + 301571 t^{20}+\ldots$} & \footnotesize{$8 t^3 + 80 t^5 + 432 t^7 + 1720 t^9 + 5640 t^{11} + 16024 t^{13} + 
 40752 t^{15} + 94888 t^{17} + 205440 t^{19}+\ldots$}  \\ \hline
 \hyperref[SO6:MQ202]{$\MQ_1^{2,0,2}$}  & \footnotesize{$1 + 14 t^2 + 119 t^4 + 806 t^6 + 4480 t^8 + 20886 t^{10} + 83778 t^{12}+\ldots$} & \footnotesize{$16 t^3 + 208 t^5 + 1568 t^7 + 8736 t^9 + 
 39552 t^{11}+\ldots$}  \\ \hline
 \hyperref[SO6:MQ202]{$\MQ_2^{2,0,2}$}  & \footnotesize{$1 + 20 t^2 + 175 t^4 + 32 t^5 + 1060 t^6 + 416 t^7 + 4994 t^8 + 
 2784 t^9 + 19432 t^{10} + 13376 t^{11} + 65340 t^{12} + 51584 t^{13} + 195360 t^{14} + 169120 t^{15} + 530141 t^{16} + 
 489760 t^{17} + 1326692 t^{18} + 1284544 t^{19} + 3098889 t^{20} + \ldots$} & \footnotesize{$2 t^2 + 16 t^3 + 40 t^4 + 208 t^5 + 348 t^6 + 1408 t^7 + 2080 t^8 + 6896 t^9 + 9640 t^{10} + 27184 t^{11} + 36784 t^{12} + 91248 t^{13} + 121040 t^{14} + 270672 t^{15} + 353936 t^{16} + 726832 t^{17} + 939242 t^{18} + 1797536 t^{19} + 2299448 t^{20} + \ldots$}  \\ \hline
 \hyperref[SO6:MQ202]{$\MQ_3^{2,0,2}$}  &\footnotesize{$1 + 11 t^2 + 68 t^4 + 313 t^6 + 1202 t^8 + 3953 t^{10} + 11453 t^{12} + 29842 t^{14} + 71275 t^{16} + 158094 t^{18} + 329343 t^{20} + \ldots$} & \footnotesize{$2 t^2 + 32 t^4 + 214 t^6 + 972 t^8 + 3472 t^{10} + 10544 t^{12} + 28260 t^{14} + 68662 t^{16} + 153948 t^{18} + 323034 t^{20} + \ldots$}  \\ \hline
 \hyperref[SO6:MQ212]{$\MQ_1^{2,1,2}$}  & \footnotesize{$1 + 20 t^2 + 8 t^3 + 275 t^4 + 296 t^5 + 3045 t^6 + 4800 t^7 + 
 27790 t^8 + 51128 t^9 + 211871 t^{10}+\ldots$} & \footnotesize{$6 t^2 + 24 t^3 + 150 t^4 + 504 t^5 + 2144 t^6 + 6528 t^7 + 
 22284 t^8 + 61992 t^9 + 182902 t^{10}+\ldots$}  \\ \hline
 \hyperref[SO6:MQ222]{$\MQ_1^{2,2,2}$}  & \footnotesize{$1 + 27 t^2 + 16 t^3 + 537 t^4 + 992 t^5 + 9266 t^6 + 23904 t^7 + 
 133582 t^8 + 383712 t^9 + 1626798 t^{10}+\ldots$} & \footnotesize{$12 t^2 + 48 t^3 + 384 t^4 + 1440 t^5 + 7700 t^6 + 28192 t^7 + 
 120740 t^8 + 417440 t^9 + 1536816 t^{10}+\ldots$}  \\ \hline
 \hyperref[SO6:MQtilde222]{$\widehat{\MQ}_1^{2,2,2}$}  & \footnotesize{$1 + 39 t^2 + 64 t^3 + 921 t^4 + 2432 t^5 + 16966 t^6 + 52096 t^7 + 
 254322 t^8 + 801152 t^9 + 3163614 t^{10}+\ldots$} & \footnotesize{not required}  \\ \hline
 \hyperref[SO6:MQ322]{$\MQ_1^{3,2,2}$}  & \footnotesize{$1 + 40 t^2 + 64 t^3 + 1104 t^4 + 3520 t^5 + 26972 t^6 + 108032 t^7 + 
 587528 t^8 + 2428480 t^9 + 11018073 t^{10}+\ldots$} & \footnotesize{$16 t^2 + 64 t^3 + 736 t^4 + 3520 t^5 + 22848 t^6 + 108032 t^7 + 
 549792 t^8 + 2428480 t^9 + 10724912 t^{10}+\ldots$}  \\ \hline
 \hyperref[SO6:MQ332]{$\MQ_1^{3,3,2}$}  & \footnotesize{$1 + 57 t^2 + 88 t^3 + 2432 t^4 + 9024 t^5 + 91715 t^6+\ldots$} & \footnotesize{$32 t^2 + 128 t^3 + 2016 t^4 + 9984 t^5 + 85856 t^6+\ldots$}  \\ \hline
 \hyperref[SO8:MQ001]{$\MQ_{1;\, \text{SO}(8)}^{0,0,1}$}  & \CBHS{1 + 2 t^2 + 2 t^4 + 4 t^6 + 2 t^8 + 2 t^{10} + t^{12}}{(1 - t^2)^2\,  (1 - t^6)^2}{1 + 4 t^2 + 9 t^4 + 20 t^6 + 37 t^8 + 60 t^{10}} & \footnotesize{not required}  \\ \hline
 \hyperref[SO8:MQ100]{$\MQ_{1;\, \text{SO}(8)}^{1,0,0}$}  & \CBHS{1 + 2 t^2 + 2 t^4 + 4 t^6 + 2 t^8 + 2 t^{10} + t^{12}}{(1 - t^2)^2\,  (1 - t^6)^2}{1 + 4 t^2 + 9 t^4 + 20 t^6 + 37 t^8 + 60 t^{10}} & \footnotesize{not required}  \\ \hline
 \hyperref[SO8:MQ002]{$\MQ_{1;\, \text{SO}(8)}^{0,0,2}$}  & \CBHS{P_{11}(t)}{(1 - t^2)^4\, (1 - t^6)^4}{1 + 11 t^2 + 60 t^4 + 235 t^6 + 745 t^8 + 2016 t^{10}} & \footnotesize{not required}  \\ \hline
 \hyperref[SO8:MQ200]{$\MQ_{1;\, \text{SO}(8)}^{2,0,0}$}  & \CBHS{P_{12}(t)}{(1 - t^2)^2\, (1 - t^4)^2\, (1 - t^6)^4}{1 + 7 t^2 + 36 t^4 + 133 t^6 + 409 t^8 + 1082 t^{10}} & \CBHS{P_{13}(t)}{(1 - t^2)^2\, (1 - t^4)^2\, (1 - t^6)^4}{4 t^2 + 24 t^4 + 102 t^6 + 336 t^8 + 934 t^{10}}  \\ \hline
 \hyperref[SO8:MQ210]{$\MQ_{1;\, \text{SO}(8)}^{2,1,0}$}  & \footnotesize{$1 + 10 t^2 + 62 t^4 + 291 t^6 + 1102 t^8 + 3556 t^{10} + 10104 t^{12} + 
 25904 t^{14} + 60965 t^{16} + 133590 t^{18} + 275450 t^{20}+\ldots$} & \footnotesize{$4 t^2 + 36 t^4 + 196 t^6 + 824 t^8 + 2840 t^{10} + 8448 t^{12} +  22392 t^{14} + 54040 t^{16} + 120684 t^{18} + 252596 t^{20}+\ldots$}  \\ \hline
 \hyperref[SO8:MQ201]{$\MQ_{1;\, \text{SO}(8)}^{2,0,1}$}  & \footnotesize{$1 + 10 t^2 + 62 t^4 + 291 t^6 + 1102 t^8 + 3556 t^{10} + 10104 t^{12} + 
 25904 t^{14} + 60965 t^{16} + 133590 t^{18} + 275450 t^{20}+\ldots$} & \footnotesize{$4 t^2 + 36 t^4 + 196 t^6 + 824 t^8 + 2840 t^{10} + 8448 t^{12} +  22392 t^{14} + 54040 t^{16} + 120684 t^{18} + 252596 t^{20}+\ldots$}  \\ \hline
 \hyperref[SO8:MQ102]{$\MQ_{1;\, \text{SO}(8)}^{1,0,2}$}  & \footnotesize{$1 + 14 t^2 + 98 t^4 + 487 t^6 + 1926 t^8 + 6396 t^{10} + 18552 t^{12} + 
 48296 t^{14} + 115005 t^{16} + 254274 t^{18} + 528046 t^{20}+\ldots$} & \footnotesize{not required}  \\ \hline
 \hyperref[SO8:MQ220]{$\MQ_{1;\, \text{SO}(8)}^{2,2,0}$}  & \footnotesize{$1 + 13 t^2 + 143 t^4 + 1106 t^6 + 6918 t^8 + 35792 t^{10} +159285 t^{12} + 623177 t^{14} + 2187539 t^{16} + 6992878 t^{18} + 
 20617582 t^{20}+\ldots$} & \footnotesize{$8 t^2 + 104 t^4 + 936 t^6 + 6200 t^8 + 33384 t^{10} + 151776 t^{12} + 
 602624 t^{14} + 2134800 t^{16} + 6868552 t^{18} + 20339496 t^{20}+\ldots$}  \\ \hline
 \hyperref[SO8:MQ220]{$\MQ_{2;\, \text{SO}(8)}^{2,2,0}$}  & \footnotesize{$1 + 6 t^2 + 30 t^4 + 110 t^6 + 339 t^8 + 900 t^{10} + 2140 t^{12} + 
 4644 t^{14} + 9365 t^{16} + 17754 t^{18} + 31962 t^{20}+\ldots$} & \footnotesize{$4 t^2 + 24 t^4 + 100 t^6 + 320 t^8 + 872 t^{10} + 2096 t^{12} + 4584 t^{14} + 9280 t^{16} + 17644 t^{18} + 31816 t^{20}+\ldots$}  \\ \hline
 \hyperref[SO8:MQ202]{$\MQ_{1;\, \text{SO}(8)}^{2,0,2}$}  & \footnotesize{$1 + 17 t^2 + 161 t^4 + 1178 t^6 + 7146 t^8 + 36564 t^{10} + 161447 t^{12} + 628953 t^{14} + 2201473 t^{16} + 7025402 t^{18} + 
 20688074 t^{20}+\ldots$} & \footnotesize{$4 t^2 + 86 t^4 + 864 t^6 + 5972 t^8 + 32612 t^{10} + 149614 t^{12} + 
 596848 t^{14} + 2120866 t^{16} + 6836028 t^{18} + 20269004 t^{20}+\ldots$}  \\ \hline
 \hyperref[SO8:MQ202]{$\MQ_{2;\, \text{SO}(8)}^{2,0,2}$}  & \footnotesize{$1 + 6 t^2 + 30 t^4 + 110 t^6 + 339 t^8 + 900 t^{10} + 2140 t^{12} + 
 4644 t^{14} + 9365 t^{16} + 17754 t^{18} + 31962 t^{20}+\ldots$} & \footnotesize{$4 t^2 + 24 t^4 + 100 t^6 + 320 t^8 + 872 t^{10} + 2096 t^{12} + 4584 t^{14} + 9280 t^{16} + 17644 t^{18} + 31816 t^{20}+\ldots$}  \\ \hline
 \hyperref[SO8:MQ112]{$\MQ_{1;\, \text{SO}(8)}^{1,1,2}$}  & \footnotesize{$1 + 17 t^2 + 16 t^3 + 151 t^4 + 248 t^5 + 1065 t^6 + 2032 t^7 + 
 6375 t^8 + 12320 t^9 + 32229 t^{10} + 61368 t^{11} + 140928 t^{12} + 
 261264 t^{13} + 545817 t^{14} + 978968 t^{15} + 1902162 t^{16} + 
 3301248 t^{17} + 6048732 t^{18} + 10175424 t^{19} + 17765022 t^{20}+\ldots$} & \footnotesize{not required}  \\ \hline
 \hyperref[SO8:MQ112]{$\MQ_{2;\, \text{SO}(8)}^{1,1,2}$}  & \CBHS{P_{14}(t)}{(1 - t^2)^4\, (1 - t^4)^4}{1 + 10 t^2 + 55 t^4 + 215 t^6 + 679 t^8 + 1831 t^{10}} & \footnotesize{not required}  \\ \hline
 \hyperref[SO8:MQ211]{$\MQ_{1;\, \text{SO}(8)}^{2,1,1}$}  & \footnotesize{$1 + 13 t^2 + 8 t^3 + 99 t^4 + 124 t^5 + 645 t^6 + 1016 t^7 + 
 3631 t^8 + 6160 t^9 + 17605 t^{10} + 30684 t^{11} + 74988 t^{12} + 
 130632 t^{13} + 285329 t^{14} + 489484 t^{15} + 982434 t^{16} + 
 1650624 t^{17} + 3098316 t^{18} + 5087712 t^{19} + 9046590 t^{20}+\ldots$} & \footnotesize{$4 t^2 + 8 t^3 + 52 t^4 + 124 t^5 + 420 t^6 + 1016 t^7 + 2744 t^8 + 
 6160 t^9 + 14624 t^{10} + 30684 t^{11} + 65940 t^{12} + 130632 t^{13} + 
 260488 t^{14} + 489484 t^{15} + 919728 t^{16} + 1650624 t^{17} + 
 2950416 t^{18} + 5087712 t^{19} + 8718432 t^{20}+\ldots$}  \\ \hline
 \hyperref[SO8:MQ211]{$\MQ_{2;\, \text{SO}(8)}^{2,1,1}$}  & \footnotesize{$1 + 6 t^2 + 31 t^4 + 111 t^6 + 351 t^8 + 927 t^{10} + 2222 t^{12} + 
 4811 t^{14} + 9745 t^{16} + 18463 t^{18} + 33309 t^{20}+\ldots$} & \footnotesize{$4 t^2 + 24 t^4 + 104 t^6 + 328 t^8 + 904 t^{10} + 2168 t^{12} + 
 4768 t^{14} + 9640 t^{16} + 18376 t^{18} + 33128 t^{20}+\ldots$}  \\ \hline
 \hyperref[SO8:MQ221]{$\MQ_{1;\, \text{SO}(8)}^{2,2,1}$}  & \footnotesize{$1 + 16 t^2 + 16 t^3 + 192 t^4 + 384 t^5 + 2037 t^6 + 5072 t^7 + 
 19123 t^8 + 49824 t^9 + 156259 t^{10}+\ldots$} & \footnotesize{$8 t^2 + 16 t^3 + 136 t^4 + 384 t^5 + 1688 t^6 + 5072 t^7 + 
 17400 t^8 + 49824 t^9 + 148624 t^{10}+\ldots$}  \\ \hline
 \hyperref[SO8:MQ221]{$\MQ_{2;\, \text{SO}(8)}^{2,2,1}$}  & \footnotesize{$1 + 15 t^2 + 16 t^3 + 175 t^4 + 368 t^5 + 1840 t^6 + 4768 t^7 + 
 16982 t^8 + 45408 t^9 + 135469 t^{10}+\ldots$} & \footnotesize{$8 t^2 + 16 t^3 + 136 t^4 + 368 t^5 + 1640 t^6 + 4768 t^7 + 
 16208 t^8 + 45408 t^9 + 132744 t^10+\ldots$}  \\ \hline
 \hyperref[SO8:MQ212]{$\MQ_{1;\, \text{SO}(8)}^{2,1,2}$}  & \footnotesize{$1 + 20 t^2 + 16 t^3 + 224 t^4 + 384 t^5 + 2189 t^6 + 5072 t^7 + 
 19815 t^8 + 49824 t^9 + 159015 t^{10}+\ldots$} & \footnotesize{$4 t^2 + 16 t^3 + 104 t^4 + 384 t^5 + 1536 t^6 + 5072 t^7 + 
 16708 t^8 + 49824 t^9 + 145868 t^{10}+\ldots$}  \\ \hline
 \hyperref[SO8:MQ212]{$\MQ_{2;\, \text{SO}(8)}^{2,1,2}$}  & \footnotesize{$1 + 19 t^2 + 16 t^3 + 211 t^4 + 368 t^5 + 2020 t^6 + 4768 t^7 + 
 17758 t^8 + 45408 t^9 + 138385 t^{10}+\ldots$} & \footnotesize{$4 t^2 + 16 t^3 + 100 t^4 + 368 t^5 + 1460 t^6 + 4768 t^7 + 
 15432 t^8 + 45408 t^9 + 129828 t^{10}+\ldots$}  \\ \hline
	\end{longtable}
\captionsetup{width=15cm}
\begin{longtable}{|c|C{12.25cm}|}
\caption{Unrefined Higgs branch Hilbert series for the orthosymplectic quivers.}
			\label{HiggsHSOSp} \\ \hline
  Quiver & Higgs branch HS: $\text{HS}_{\mathcal{H}}(t)$  \\ \hline
	\hyperref[SO6:MQ110]{$\MQ_1^{1,1,0}$} & \HBHS{(1-t)\,Q_1(t)}{(1-t^2)^4\,(1-t^3)^3}{1 + 9 t^2 + 6 t^3 + 36 t^4 + 36 t^5 + 112 t^6 + 120 t^7 + 285 t^8 \\ + 
 320 t^9+ 621 t^{10}} \\ \hline
 \hyperref[SO6:MQ200]{$\MQ_1^{2,0,0}$} & \HBHS{1 + 9 t^2 + 9 t^4 + t^6}{(1-t^2)^6}{1 + 15 t^2 + 84 t^4 + 300 t^6 + 825 t^8 + 1911 t^{10}} \\ \hline
 \hyperref[SO6:MQ200]{$\MQ_2^{2,0,0}$} & \HBHS{1 + t^2}{(1-t^2)^2}{1 + 3 t^2 + 5 t^4 + 7 t^6 + 9 t^8 + 11 t^{10}} \\ \hline
\hyperref[SO6:MQ210]{$\MQ_1^{2,1,0}$} & \HBHS{Q_2(t)}{(1-t^2)^3\,(1-t^4)^3}{1 + 9 t^2 + 42 t^4 + 136 t^6 + 357 t^8 + 801 t^{10}} \\ \hline
\hyperref[SO6:MQ220]{$\MQ_1^{2,2,0}$} & \HBHS{Q_3(t)}{(1-t^2)\,(1-t^3)^2\,(1-t^4)\,(1-t^5)^2}{1 + 4 t^2 + 4 t^3 + 11 t^4 + 16 t^5 + 31 t^6 + 44 t^7\\ + 72 t^8 + 
 104 t^9 + 155 t^{10}} \\ \hline
 \hyperref[SO6:MQ220]{$\MQ_2^{2,2,0}$}  & \HBHS{1 + 9 t^2 + 9 t^4 + t^6}{(1-t^2)^6}{1 + 15 t^2 + 84 t^4 + 300 t^6 + 825 t^8 + 1911 t^{10}}  \\ \hline
\hyperref[SO6:MQtilde220]{$\widehat{\MQ}_1^{2,2,0}$} & \HBHS{Q_3(t)}{(1-t^2)\,(1-t^3)^2\,(1-t^4)\,(1-t^5)^2}{1 + 4 t^2 + 4 t^3 + 11 t^4 + 16 t^5 + 31 t^6 + 44 t^7\\ + 72 t^8 + 
 104 t^9 + 155 t^{10}} \\ \hline
\hyperref[SO6:MQtilde220]{$\widehat{\MQ}_2^{2,2,0}$} & \HBHS{1 + 9 t^2 + 9 t^4 + t^6}{(1-t^2)^6}{1 + 15 t^2 + 84 t^4 + 300 t^6 + 825 t^8 + 1911 t^{10}} \\\hline
\hyperref[SO6:MQ320]{$\MQ_1^{3,2,0}$} & \HBHS{Q_4(t)}{(1-t^2)\,(1-t^4)^3\,(1-t^6)^2}{1 + 4 t^2 + 14 t^4 + 37 t^6 + 86 t^8 + 176 t^{10}} \\ \hline
\hyperref[SO6:MQ320]{$\MQ_2^{3,2,0}$} & \HBHS{Q_5(t)}{(1-t^2)^3\,(1-t^6)^3}{1 + 9 t^2 + 36 t^4 + 106 t^6 + 261 t^8 + 561 t^{10}} \\ \hline
 \hyperref[SO6:MQ330]{$\MQ_1^{3,3,0}$} & \HBHS{(1-t)Q_6(t)}{(1-t^2)\,(1-t^3)\,(1-t^4)^2\,(1-t^5)\,(1-t^6)\,(1-t^7)}{1 + t^2 + 2 t^3 + 3 t^4 + 4 t^5 + 8 t^6\\ + 10 t^7 + 15 t^8 + 20 t^9 + 
 30 t^{10}} \\\hline
\hyperref[SO6:MQ330]{$\MQ_2^{3,3,0}$} & \HBHS{Q_7(t)}{(1-t^2)\,(1-t^5)^2\,(1-t^7)^2\,(1-t^8)}{1 + 4 t^2 + 10 t^4 + 4 t^5 + 20 t^6\\ + 16 t^7 + 36 t^8 + 40 t^9 + 
 67 t^{10}} \\ \hline
 \hyperref[SO6:MQ440]{$\MQ_1^{4,4,0}$}  & \footnotesize{$1 + t^4 + 2 t^6 + 5 t^8 + 5 t^{10} + 12 t^{12}+\ldots$}  \\ \hline
 \hyperref[SO6:MQ201]{$\MQ_1^{2,0,1}$}  & \HBHS{(1-t)Q_8(t)}{(1-t^2)^2\,(1-t^4)^2\,(1-t^3)^3}{1 + 5 t^2 + 6 t^3 + 18 t^4 + 26 t^5 + 58 t^6 \\+ 78 t^7 + 149 t^8 + 
 198 t^9 + 325 t^{10}}  \\ \hline
  \hyperref[SO6:MQ201]{$\MQ_2^{2,0,1}$}  & \HBHS{(1+t^2)^2}{(1-t^2)^4}{1 + 6 t^2 + 19 t^4 + 44 t^6 + 85 t^8 + 146 t^{10}}  \\ \hline
  \hyperref[SO6:MQ201]{$\MQ_3^{2,0,1}$}  & \HBHS{1 + 5 t^2 + 5 t^4 + t^6}{(1-t^2)^6}{1 + 11 t^2 + 56 t^4 + 192 t^6 + 517 t^8 + 1183 t^{10}}  \\ \hline
  \hyperref[SO6:MQ111]{$\MQ_1^{1,1,1}$}  & \HBHS{Q_9(t)}{(1-t)\,(1-t^2)\,(1-t^3)^3\,(1-t^4)}{1 + 5 t^2 + 8 t^3 + 16 t^4 + 32 t^5\\ + 58 t^6 + 88 t^7 + 151 t^8 + 
 224 t^9 + 329 t^{10}}  \\ \hline
 \hyperref[SO6:MQ111]{$\MQ_2^{1,1,1}$}  & \HBHS{1 + 2 t^2 + 2 t^3 + 2 t^4 + t^6}{(1-t^2)^2\,(1-t^3)^2}{1 + 4 t^2 + 4 t^3 + 9 t^4 + 12 t^5\\ + 22 t^6 + 24 t^7 + 41 t^8 + 
 48 t^9 + 66 t^{10}}  \\ \hline
  \hyperref[SO6:MQ211]{$\MQ_1^{2,1,1}$}  & \HBHS{(1-t)\,Q_{10}(t)}{(1-t^2)^2\,(1-t^4)^2\,(1-t^5)^3}{1 + 5 t^2 + 18 t^4 + 6 t^5 + 46 t^6\\ + 26 t^7 + 101 t^8 + 78 t^9 + 
 205 t^{10}}  \\ \hline
  \hyperref[SO6:MQ211]{$\MQ_2^{2,1,1}$}  & \HBHS{1 + 2 t^2 + 2 t^4 + t^6}{(1-t^2)^6}{1 + 8 t^2 + 35 t^4 + 111 t^6 + 286 t^8 + 637 t^{10}}  \\ \hline
  \hyperref[SO6:MQ221]{$\MQ_1^{2,2,1}$} & \HBHS{Q_{11}(t)}{(1-t^4)^2\,(1-t^5)\,(1-t^6)^2\,(1-t^7)}{1 + t^2 + 2 t^3 + 4 t^4 + 4 t^5 + 9 t^6 + 12 t^7 \\+ 20 t^8 + 26 t^9 + 
 39 t^{10}} \\\hline
 \hyperref[SO6:MQ221]{$\MQ_2^{2,2,1}$} & \HBHS{Q_{16}(t)}{(1 - t^2)^2\, (1 - t^5)^2\, (1 - t^6)^2}{1 + 4 t^2 + 10 t^4 + 4 t^5 + 23 t^6 + 16 t^7 + 46 t^8 + 40 t^9 + 
 88 t^{10}} \\\hline
  \hyperref[SO6:MQhat221]{$\widehat{\MQ}_1^{2,2,1}$} & \footnotesize{$1 + t^2 + 4 t^5 + 4 t^5 + 9t^6 + 12t^7 + 20t^8 +\ldots$} \\\hline
  \hyperref[SO6:MQhat221]{$\widehat{\MQ}_2^{2,2,1}$} & \HBHS{Q_{16}(t)}{(1 - t^2)^2\, (1 - t^5)^2\, (1 - t^6)^2}{1 + 4 t^2 + 10 t^4 + 4 t^5 + 23 t^6 + 16 t^7 + 46 t^8 + 40 t^9 + 
 88 t^{10}} \\\hline
  \hyperref[SO6:MQ321]{$\MQ_1^{3,2,1}$}  & \footnotesize{$1 + t^2 + 2 t^4 + 2 t^5 + 4 t^6 + 4 t^7 + 8 t^8 + 8 t^9 + 13 t^{10} + \ldots$} \\ \hline
  \hyperref[SO6:MQ331]{$\MQ_1^{3,3,1}$}  & \footnotesize{$1 + t^4 + 2 t^6 + 5 t^8+\ldots$} \\ \hline
  \hyperref[SO6:MQ441]{$\MQ_1^{4,4,1}$}  & \footnotesize{$1+\mathcal{O}(t^7)$} \\ \hline
  \hyperref[SO6:MQ112]{$\MQ_1^{1,1,2}$}  & \HBHS{Q_{12}(t)}{(1 - t)\, (1 - t^3)\, (1 - t^4)\, (1 - t^5)^2\, (1 - t^6)}{1 + t^2 + 2 t^3 + 4 t^4 + 6 t^5\\ + 10 t^6 + 14 t^7 + 22 t^8 + 32 t^9 + 
 46 t^{10}} \\ \hline
 \hyperref[SO6:MQ112]{$\MQ_2^{1,1,2}$}  & \HBHS{Q_{13}(t)}{(1 - t)\, (1 - t^2)^2\, (1 - t^3)\, (1 - t^5)^2}{1 + 5 t^2 + 2 t^3 + 14 t^4 + 14 t^5\\ + 32 t^6 + 44 t^7 + 73 t^8 + 
 102 t^9 + 157 t^{10}} \\ \hline
 \hyperref[SO6:MQ112]{$\MQ_3^{1,1,2}$}  &  \footnotesize{$1+3t^2+14t^4+34t^6+89t^8+\cdots$}  \\ \hline
 \hyperref[SO6:MQtilde112]{$\widehat{\MQ}_1^{1,1,2}$}  & \HBHS{Q_{12}(t)}{(1 - t)\, (1 - t^3)\, (1 - t^4)\, (1 - t^5)^2\, (1 - t^6)}{1 + t^2 + 2 t^3 + 4 t^4 + 6 t^5\\ + 10 t^6 + 14 t^7 + 22 t^8 + 32 t^9 + 
 46 t^{10}} \\ \hline
 \hyperref[SO6:MQtilde112]{$\widehat{\MQ}_2^{1,1,2}$}   & \HBHS{1 - t + 3 t^2 - 2 t^3 + 4 t^4 + 2 t^6 + 4 t^8 - 2 t^9 + 
 3 t^{10} - t^{11} + t^{12}}{(1 - t)\, (1 - t^2)^2\, (1 - t^3)\, (1 - t^5)^2}{1 + 5 t^2 + 2 t^3 + 14 t^4 + 14 t^5 + 32 t^6 + 44 t^7 + 73 t^8 + 
 102 t^9 + 157 t^{10}} \\ \hline
 \hyperref[SO6:MQtilde112]{$\widehat{\MQ}_3^{1,1,2}$}  & \HBHS{1 + 5 t^4 + 5 t^8 + t^{12}}{(1 - t^2)^3\, (1 - t^4)^3}{1 + 3 t^2 + 14 t^4 + 34 t^6 + 89 t^8 + 179 t^{10}} \\ \hline
 \hyperref[SO6:MQ212]{$\MQ_1^{2,1,2}$}  &\footnotesize{$1+t^2+3t^4+2t^5+4t^6+6t^7+8t^8+\ldots$}   \\ \hline
 \hyperref[SO6:MQ222]{$\MQ_1^{2,2,2}$}  & \footnotesize{$1+t^4+2t^6+5t^8+\ldots$}   \\ \hline
 \hyperref[SO6:MQtilde222]{$\widehat{\MQ}_1^{2,2,2}$} & \footnotesize{$1+t^4+2t^6+5t^8+\ldots$}   \\ \hline
 \hyperref[SO6:MQ322]{$\MQ_1^{3,2,2}$}  & \footnotesize{$1+t^6+\ldots$} \\ \hline
 \hyperref[SO6:MQ332]{$\MQ_1^{3,3,2}$}  & \footnotesize{$1 + 2 t^2 + 8 t^4+\ldots$} \\ \hline
 \hyperref[SO8:MQ001]{$\MQ_{1;\, \text{SO}(8)}^{0,0,1}$}  & \HBHS{(1-t)\,Q_{13}(t)}{(1 - t^2)^4\, (1 - t^3)^4\, (1 - t^5)}{1 + 10 t^2 + 18 t^3 + 52 t^4 + 116 t^5 + 250 t^6 + 454 t^7 + 
 889 t^8 + 1490 t^9 + 2538 t^{10}}  \\ \hline
 \hyperref[SO8:MQ100]{$\MQ_{1;\, \text{SO}(8)}^{1,0,0}$}  & \HBHS{(1-t)\,Q_{13}(t)}{(1 - t^2)^4\, (1 - t^3)^4\, (1 - t^5)}{1 + 10 t^2 + 18 t^3 + 52 t^4 + 116 t^5 + 250 t^6 + 454 t^7 + 
 889 t^8 + 1490 t^9 + 2538 t^{10}}  \\ \hline
 \hyperref[SO8:MQ002]{$\MQ_{1;\, \text{SO}(8)}^{0,0,2}$}  & \HBHS{1 + 6 t^2 + 27 t^4 + 48 t^6 + 84 t^8 + 86 t^{10} + 84 t^{12} + 48 t^{14} + 
 27 t^{16} + 6 t^{18} + t^{20}}{(1 - t^2)^3\, (1 - t^4)^4\, (1 - t^6)}{1 + 9 t^2 + 55 t^4 + 212 t^6 + 688 t^8 + 1852 t^{10}}  \\ \hline
 \hyperref[SO8:MQ200]{$\MQ_{1;\, \text{SO}(8)}^{2,0,0}$}  & \HBHS{1 + 6 t^2 + 27 t^4 + 48 t^6 + 84 t^8 + 86 t^{10} + 84 t^{12} + 48 t^{14} + 
 27 t^{16} + 6 t^{18} + t^{20}}{(1 - t^2)^3\, (1 - t^4)^4\, (1 - t^6)}{1 + 9 t^2 + 55 t^4 + 212 t^6 + 688 t^8 + 1852 t^{10}}  \\ \hline
 \hyperref[SO8:MQ210]{$\MQ_{1;\, \text{SO}(8)}^{2,1,0}$}  & \footnotesize{$1 + 5 t^2 + 4 t^3 + 25 t^4 + 22 t^5 + 86 t^6 + 86 t^7 + 254 t^8 + 
 270 t^9 + 648 t^{10} + 716 t^{11} + 1499 t^{12} + 1686 t^{13} + 3177 t^{14}+\ldots$}  \\ \hline
 \hyperref[SO8:MQ201]{$\MQ_{1;\, \text{SO}(8)}^{2,0,1}$}  & \HBHS{(1-t)\,Q_{14}(t)}{(1 - t^2)^3\, (1 - t^3)^2\, (1 - t^4)^2\, (1 - t^5)\, (1 - t^6)}{1 + 5 t^2 + 4 t^3 + 25 t^4 + 22 t^5 + 86 t^6 + 86 t^7 + 254 t^8 + 
 270 t^9 + 648 t^{10}}  \\ \hline
 \hyperref[SO8:MQ102]{$\MQ_{1;\, \text{SO}(8)}^{1,0,2}$}  & \HBHS{(1-t)\,Q_{14}(t)}{(1 - t^2)^3\, (1 - t^3)^2\, (1 - t^4)^2\, (1 - t^5)\, (1 - t^6)}{1 + 5 t^2 + 4 t^3 + 25 t^4 + 22 t^5 + 86 t^6 + 86 t^7 + 254 t^8 + 
 270 t^9 + 648 t^{10}}  \\ \hline
 \hyperref[SO8:MQ220]{$\MQ_{1;\, \text{SO}(8)}^{2,2,0}$}  & \footnotesize{$1 + 4 t^2 + 11 t^4 + 8 t^5 + 31 t^6 + 32 t^7 + 70 t^8 + 88 t^9 + 
 170 t^{10}+\ldots$} \\ \hline
 \hyperref[SO8:MQ220]{$\MQ_{2;\, \text{SO}(8)}^{2,2,0}$}  & \HBHS{1 + 2 t^2 + 2 t^4 + 2 t^6 + t^8}{(1 - t^2)^8}{1 + 10 t^2 + 54 t^4 + 210 t^6 + 659 t^8 + 1772 t^{10}}  \\ \hline
 \hyperref[SO8:MQ202]{$\MQ_{1;\, \text{SO}(8)}^{2,0,2}$}  & \footnotesize{$1 + 4 t^2 + 11 t^4 + 8 t^5 + 31 t^6 + 32 t^7 + 70 t^8 + 88 t^9 + 
 170 t^{10}+\ldots$}  \\ \hline
 \hyperref[SO8:MQ202]{$\MQ_{2;\, \text{SO}(8)}^{2,0,2}$}  & \HBHS{1 + 2 t^2 + 2 t^4 + 2 t^6 + t^8}{(1 - t^2)^8}{1 + 10 t^2 + 54 t^4 + 210 t^6 + 659 t^8 + 1772 t^{10}}  \\ \hline
 \hyperref[SO8:MQ112]{$\MQ_{1;\, \text{SO}(8)}^{1,1,2}$}  & \footnotesize{$1+4t^2+10t^4+8t^5+28t^6+28t^7+63t^8+\cdots$}   \\ \hline
 \hyperref[SO8:MQ112]{$\MQ_{2;\, \text{SO}(8)}^{1,1,2}$}  & \HBHS{1 + 6 t^2 + 17 t^4 + 27 t^6 + 32 t^8 + 27 t^{10} + 17 t^{12} + 
 6 t^{14} + t^{16}}{(1 - t^2)^4\, (1 - t^4)^4}{1 + 10 t^2 + 55 t^4 + 215 t^6 + 679 t^8 + 1831 t^{10}}  \\ \hline
 \hyperref[SO8:MQ211]{$\MQ_{1;\, \text{SO}(8)}^{2,1,1}$}  & \HBHS{(1 - t)\, Q_{15}(t)}{(1 - t^3)^2\, (1 - t^4)\, (1 - t^5)^3\, (1 - t^6)^2\, (1 - t^8)}{1 + 4 t^2 + 10 t^4 + 8 t^5 + 28 t^6 + 28 t^7 + 63 t^8 + 76 t^9 + 148 t^{10}} \\ \hline
 \hyperref[SO8:MQ211]{$\MQ_{2;\, \text{SO}(8)}^{2,1,1}$}  & \HBHS{1 + 6 t^2 + 17 t^4 + 27 t^6 + 32 t^8 + 27 t^{10} + 17 t^{12} + 
 6 t^{14} + t^{16}}{(1 - t^2)^4\, (1 - t^4)^4}{1 + 10 t^2 + 55 t^4 + 215 t^6 + 679 t^8 + 1831 t^{10}}  \\ \hline
 \hyperref[SO8:MQ221]{$\MQ_{1;\, \text{SO}(8)}^{2,2,1}$}  & \footnotesize{$1+t^2+2t^3+2t^4+2t^5+6t^6+8t^7+13t^8+\ldots$}   \\ \hline
 \hyperref[SO8:MQ221]{$\MQ_{2;\, \text{SO}(8)}^{2,2,1}$}  & \footnotesize{$1+3t^2+8t^4+2t^5+16t^6+10t^7+32t^8+\ldots$}  \\ \hline
 \hyperref[SO8:MQ212]{$\MQ_{1;\, \text{SO}(8)}^{2,1,2}$}  & \footnotesize{$1+t^2+2t^3+2t^4+2t^5+6t^6+8t^7+13t^8+\ldots$}   \\ \hline
 \hyperref[SO8:MQ212]{$\MQ_{2;\, \text{SO}(8)}^{2,1,2}$}  & \footnotesize{$1+3t^2+8t^4+2t^5+16t^6+10t^7+32t^8+\ldots$}  \\ \hline
	\end{longtable}
\captionsetup{width=15cm}
\begin{longtable}{|c|C{14cm}|} 
\caption{Palindromic polynomials appearing in the main sections.}
\label{PalPol} \\ \hline
  Label & Palindromic Polynomial \\ \hline
  $P_1(t)$ & \footnotesize{$1 + 2 t^2 + 10 t^4 + 10 t^6 + 16 t^8 + 10 t^{10} + 10 t^{12} + 
 2 t^{14} + t^{16}$} \\ \hline
  $P_2(t)$ & \footnotesize{$1 + 3 t^2 + 38 t^4 + 69 t^6 + 225 t^8 + 240 t^{10} + 372 t^{12} + 
 240 t^{14} + 225 t^{16} + 69 t^{18} + 38 t^{20} + 3 t^{22} + t^{24}$} \\ \hline
  $P_3(t)$ & \footnotesize{$8 t^2 + 20 t^4 + 112 t^6 + 156 t^8 + 328 t^{10} + 276 t^{12} + 328 t^{14} + 156 t^{16} + 112 t^{18} + 20 t^{20} + 8 t^{22}$} \\ \hline
  $P_4(t)$ & \footnotesize{$1 + 11 t^2 + 57 t^4 + 170 t^6 + 324 t^8 + 398 t^{10} + 324 t^{12} + 
 170 t^{14} + 57 t^{16} + 11 t^{18} + t^{20}$} \\ \hline
  $P_5(t)$ & \footnotesize{$1 + 12 t^2 + 63 t^4 + 204 t^6 + 550 t^8 + 1094 t^{10} + 1906 t^{12} + 
 2708 t^{14} + 3432 t^{16} + 3596 t^{18} + 3432 t^{20} + 2708 t^{22} + 
 1906 t^{24} + 1094 t^{26} + 550 t^{28} + 204 t^{30} + 63 t^{32} + 
 12 t^{34} + t^{36}$} \\ \hline
  $P_6(t)$ & \footnotesize{$8 t^2 + 56 t^4 + 216 t^6 + 536 t^8 + 1136 t^{10} + 1888 t^{12} + 
 2752 t^{14} + 3344 t^{16} + 3664 t^{18} + 3344 t^{20} + 2752 t^{22} + 
 1888 t^{24} + 1136 t^{26} + 536 t^{28} + 216 t^{30} + 56 t^{32} + 8 t^{34}$} \\ \hline
  $P_7(t)$ & \footnotesize{$1 + 3 t^2 + 8 t^4 + 21 t^6 + 33 t^8 + 34 t^{10} + 33 t^{12} + 21 t^{14} + 
 8 t^{16} + 3 t^{18} + t^{20}$} \\ \hline
  $P_8(t)$ & \footnotesize{$3 + 10 t^2 + 20 t^4 + 31 t^6 + 38 t^8 + 31 t^{10} + 20 t^{12} + 10 t^{14} + 
 3 t^{16}$} \\ \hline
  $P_9(t)$ & \footnotesize{$1 + 2 t^2 + 7 t^4 + 12 t^6 + 22 t^8 + 16 t^{10} + 22 t^{12} + 12 t^{14} + 
 7 t^{16} + 2 t^{18} + t^{20}$} \\ \hline
  $P_{10}(t)$ & \footnotesize{$2 t^2 + 4 t^4 + 2 t^6 + 10 t^8 + 2 t^{10} + 4 t^{12} + 2 t^{14}$} \\ \hline
  $P_{11}(t)$ & \footnotesize{$1 + 7 t^2 + 22 t^4 + 53 t^6 + 94 t^8 + 129 t^{10} + 148 t^{12} + 129 t^{14} + 94 t^{16} + 53 t^{18} + 22 t^{20} + 7 t^{22} + t^{24}$} \\ \hline
  $P_{12}(t)$ & \footnotesize{$1 + 5 t^2 + 21 t^4 + 54 t^6 + 114 t^8 + 182 t^{10} + 248 t^{12} + 270 t^{14} + 248 t^{16} + 182 t^{18} + 114 t^{20} + 54 t^{22} + 21 t^{24} + 5 t^{26} + t^{28}$} \\ \hline
  $P_{13}(t)$ & \footnotesize{$4 t^2 + 16 t^4 + 50 t^6 + 108 t^8 + 188 t^{10} + 252 t^{12} + 284 t^{14} + 252 t^{16} + 188 t^{18} + 108 t^{20} + 50 t^{22} + 16 t^{24} + 4 t^{26}$} \\ \hline
  $P_{14}(t)$ & \footnotesize{$1 + 6 t^2 + 17 t^4 + 27 t^6 + 32 t^8 + 27 t^{10} + 17 t^{12} + 
 6 t^{14} + t^{16}$} \\ \hline
$Q_1(t)$ & \footnotesize{$1 + t + 6 t^2 + 9 t^3 + 15 t^4 + 12 t^5 + 15 t^6 + 9 t^7 + 
 6 t^8 + t^9 + t^{10}$} \\ \hline
 $Q_2(t)$ & \footnotesize{$1 + 6 t^2 + 15 t^4 + 18 t^6 + 15 t^8 + 6 t^{10} + t^{12}$} \\ \hline
 $Q_3(t)$ & \footnotesize{$1 + 3 t^2 + 2 t^3 + 6 t^4 + 4 t^5 + 10 t^6 + 6 t^7 + 9 t^8 + 6 t^9 + 
 10 t^{10} + 4 t^{11} + 6 t^{12} + 2 t^{13} + 3 t^{14} + t^{16}$} \\ \hline
 $Q_4(t)$ & \footnotesize{$1 + 3 t^2 + 7 t^4 + 12 t^6 + 16 t^8 + 16 t^{10} + 16 t^{12} + 12 t^{14} + 7 t^{16} + 3 t^{18} + t^{20}$} \\ \hline
 $Q_5(t)$ & \footnotesize{$1 + 6 t^2 + 12 t^4 + 21 t^6 + 24 t^8 + 24 t^{10} + 21 t^{12} + 12 t^{14} + 
 6 t^{16} + t^{18}$} \\ \hline
 $Q_6(t)$ & \footnotesize{$1 + t + t^2 + 2 t^3 + 2 t^4 + 3 t^5 + 5 t^6 + 6 t^7 + 7 t^8 + 7 t^9 + 7 t^{10} + 7 t^{11} + 8 t^{12} + 7 t^{13} + 7 t^{14} + 7 t^{15} + 7 t^{16} + 6 t^{17} + 5 t^{18} + 3 t^{19} + 2 t^{20} + 2 t^{21} + t^{22} + t^{23} + t^{24}$} \\ \hline
 $Q_7(t)$ & \footnotesize{$1 + 3 t^2 + 6 t^4 + 2 t^5 + 10 t^6 + 4 t^7 + 15 t^8 + 6 t^9 + 
 21 t^{10} + 8 t^{11} + 22 t^{12} + 10 t^{13} + 25 t^{14} + 10 t^{15} + 22 t^{16} + 
 8 t^{17} + 21 t^{18} + 6 t^{19} + 15 t^{20} + 4 t^{21} + 10 t^{22} + 2 t^{23} + 
 6 t^{24} + 3 t^{26} + t^{28}$} \\ \hline
 $Q_8(t)$ & \footnotesize{$1 + t + 4 t^2 + 7 t^3 + 14 t^4 + 19 t^5 + 25 t^6 + 24 t^7 + 25 t^8 + 
 19 t^9 + 14 t^{10} + 7 t^{11} + 4 t^{12} + t^{13} + t^{14}$} \\ \hline
 $Q_9(t)$ & \footnotesize{$1 - t + 4 t^2 + t^3 + 5 t^4 + 2 t^5 + 5 t^6 + t^7 + 4 t^8 - t^9 + t^{10}$} \\ \hline
 $Q_{10}(t)$ & \footnotesize{$1 + t + 4 t^2 + 4 t^3 + 11 t^4 + 14 t^5 + 23 t^6 + 28 t^7 + 38 t^8 + 
 37 t^9 + 44 t^{10} + 37 t^{11} + 38 t^{12} + 28 t^{13} + 23 t^{14} + 14 t^{15} + 
 11 t^{16} + 4 t^{17} + 4 t^{18} + t^{19} + t^{20}$} \\ \hline
 $Q_{11}(t)$ & \footnotesize{$1 + t^2 + 2 t^3 + 2 t^4 + 3 t^5 + 5 t^6 + 6 t^7 + 9 t^8 + 11 t^9 + 
 12 t^{10} + 13 t^{11} + 16 t^{12} + 12 t^{13} + 16 t^{14} + 13 t^{15} + 
 12 t^{16} + 11 t^{17} + 9 t^{18} + 6 t^{19} + 5 t^{20} + 3 t^{21} + 2 t^{22} + 
 2 t^{23} + t^{24} + t^{26}$} \\ \hline
 $Q_{12}(t)$ & \footnotesize{$1 - t + t^2 + 2 t^4 + 3 t^6 + t^7 + 2 t^8 + t^9 + 2 t^{10} + t^{11} + 
 3 t^{12} + 2 t^{14} + t^{16} - t^{17} + t^{18}$} \\ \hline
  $Q_{13}(t)$ & \footnotesize{$1 + t + 7 t^2 + 21 t^3 + 39 t^4 + 58 t^5 + 90 t^6 + 110 t^7 + 
 118 t^8 + 110 t^9 + 90 t^{10} + 58 t^{11} + 39 t^{12} + 21 t^{13} + 
 7 t^{14} + t^{15} + t^{16}$} \\ \hline
 $Q_{14}(t)$ & \footnotesize{$1 + t + 3 t^2 + 5 t^3 + 16 t^4 + 21 t^5 + 34 t^6 + 34 t^7 + 53 t^8 + 
 56 t^9 + 73 t^{10} + 64 t^{11} + 73 t^{12} + 56 t^{13} + 53 t^{14} + 34 t^{15} + 
 34 t^{16} + 21 t^{17} + 16 t^{18} + 5 t^{19} + 3 t^{20} + t^{21} + t^{22}$} \\ \hline
 $Q_{15}(t)$ & \footnotesize{$1 + t + 5 t^2 + 3 t^3 + 12 t^4 + 9 t^5 + 32 t^6 + 30 t^7 + 68 t^8 + 
 65 t^9 + 119 t^{10} + 114 t^{11} + 187 t^{12} + 181 t^{13} + 268 t^{14} + 
 248 t^{15} + 330 t^{16} + 287 t^{17} + 354 t^{18} + 287 t^{19} + 330 t^{20} + 
 248 t^{21} + 268 t^{22} + 181 t^{23} + 187 t^{24} + 114 t^{25} + 119 t^{26} + 
 65 t^{27} + 68 t^{28} + 30 t^{29} + 32 t^{30} + 9 t^{31} + 12 t^{32} + 3 t^{33} + 
 5 t^{34} + t^{35} + t^{36}$} \\ \hline
 $Q_{16}(t)$ & \footnotesize{$1 + 2 t^2 + 3 t^4 + 2 t^5 + 5 t^6 + 4 t^7 + 6 t^8 + 6 t^9 + 6 t^{10} + 
 6 t^{11} + 6 t^{12} + 4 t^{13} + 5 t^{14} + 2 t^{15} + 3 t^{16} + 2 t^{18} + t^{20}$} \\ \hline
\end{longtable}

\bibliographystyle{JHEP}

\bibliography{ref}
\end{document}